\documentclass[ba]{imsart}

\usepackage[commenters={JH}]{shortex}
\usepackage{enumitem}
\usepackage{listings}
\RequirePackage[numbers,sort&compress]{natbib}
\RequirePackage{cleveref}

\startlocaldefs
\newcommand{\Wmin}{W_{\min}}
\newcommand{\ESS}{\mathrm{ESS}}
\newcommand{\MCSE}{\mathrm{MCSE}}
\endlocaldefs

\begin{document}

\begin{frontmatter}
\title{Large-scale empirical tuning and comparison of\\default optimizers for variational inference}
%\title{A sample article title with some additional note\thanksref{t1}}
\runtitle{Default optimizers for variational inference}
%\thankstext{T1}{A sample additional note to the title.}

\begin{aug}
%%%%%%%%%%%%%%%%%%%%%%%%%%%%%%%%%%%%%%%%%%%%%%%
%% Only one address is permitted per author. %%
%% Only division, organization and e-mail is %%
%% included in the address.                  %%
%% Additional information can be included in %%
\author[UBC]{\fnms{Trevor}~\snm{Campbell}\ead[label=e1]{trevor@stat.ubc.ca}},
\author[MS,DS]{\fnms{Jonathan}~\snm{H.~Huggins}\ead[label=e2]{huggins@bu.edu}},
\author[UP]{\fnms{Kyurae}~\snm{Kim}\ead[label=e4]{kyrkim@seas.upenn.edu}},
\author[UBC]{\fnms{Charles}~\snm{C.~Margossian}\ead[label=e3]{charles.margossian@stat.ubc.ca}}
%\author[B]{\fnms{Second}~\snm{Author}\ead[label=e2]{second@somewhere.com}\orcid{0000-0000-0000-0000}}
%\and
%\author[B]{\fnms{Third}~\snm{Author}\ead[label=e3]{third@somewhere.com}}
%%%%%%%%%%%%%%%%%%%%%%%%%%%%%%%%%%%%%%%%%%%%%%
%% Addresses                                %%
%%%%%%%%%%%%%%%%%%%%%%%%%%%%%%%%%%%%%%%%%%%%%%
\address[A]{Authors listed in alphabetical order.}
\address[UBC]{Department of Statistics, UBC\printead[presep={,\ }]{e1,e3}}
\address[MS]{Department of Mathematics \& Statistics, Boston University\printead[presep={,\ }]{e2}}
\address[DS]{Faculty of Computing \& Statistics, Boston University}
\address[UP]{Department of Computer and Information Science, UPenn\printead[presep={,\ }]{e4}}

\runauthor{T.~Campbell, J.~H.~Huggins, K.~Kim, C.~C.~Margossian}
\end{aug}

\begin{abstract}
Black-box variational inference (BBVI) is a methodology for posterior approximation that relies on stochastic optimization. 
In practice, the stochastic optimizers underpinning BBVI generally require extensive problem-specific tuning,
which undermines its promise as a truly ``black box'' inference algorithm.
However,  over the past decade, many new adaptive stochastic optimization algorithms
have been developed that reduce or remove entirely the need for tuning.
In this work, we investigate this new collection of adaptive methods in the context of BBVI, with the goal of establishing
the current state of the art in tuning-free optimization-based inference.
In particular, we present a large-scale empirical evaluation
of 56 stochastic gradient-based optimization algorithms applied to 1092 Bayesian inference optimization problems, involving
over 550,000 individual optimization runs and 15 core-years of compute. 
The optimization algorithms we evaluate are chosen to represent a wide spectrum of recent approaches % to adaptive optimization,
and the benchmark problems are chosen to span a range of difficulty, with posterior target dimension $1$--$10^4$,
condition number $1$--$10^8$, and a range of variational families. % from diagonal Gaussians to neural-network-based variational autoencoders.
Our results show that no single method dominates, but running a selection of 5
algorithms suffices to reliably get close to the best-possible observed performance.  We
thus provide a strong baseline for applications where expert tuning is not
possible and for comparison when developing new stochastic optimization
algorithms. 
\end{abstract}



\end{frontmatter}

\section{Introduction}

Variational inference (VI) \cite{Blei17,Jordan99,Hinton93,Peterson89} is an
optimization-based approach to posterior approximation, where, given a target posterior
$p$, we search for the optimal approximation to $p$ in some family of
distributions $Q$. For both statistical~\cite{Dhaka21} and computational
reasons~\cite{Geffner21a,Geffner21b}, the objective is typically
the exclusive Kullback-Leibler (KL) divergence
\[
\operatorname{KL}\lt(q, p\rt) = \int \log \lt( \frac{q(z)}{p(z)} \rt) q(\mathrm{d} z),
\]
such that the approximation $q_*$ is found by solving 
\[
  q_* = \text{arg\,min}_{q \in \mathcal Q} \; \mathrm{KL}(q, p).
  \label{eq:variational_problem}
\]
This is equivalent to minimizing the negative evidence lower bound (ELBO) \cite{Jordan99},
which is the KL divergence up to the unknown log-normalizing constant of the target $p$.

Among various algorithms for solving \cref{eq:variational_problem}, black-box
variational inference (BBVI) \cite{Titsias14,Ranganath14,Wingate13,Kucukelbir17} promises general,
automated Bayesian inference by leveraging stochastic gradient-based optimization
algorithms \cite{Robbins51,Bottou99,Bottou18},
which require only pointwise evaluation of
the unnormalized target log-density and gradient \cite{Titsias14,Ranganath14,Wingate13,Kucukelbir17,Rezende14,Kingma14-vae,Graves11}. (See also~\cite{Mohamed20}.)
As such, BBVI relieves practitioners from the
various modeling constraints imposed by classical variational inference methods \cite{Jordan99,Peterson89,Minka01} such as
coordinate-ascent VI~\cite{Saul96} and expectation propagation~\cite{Minka01}.
But despite its generality and wide applicability, BBVI has fallen short on delivering a fully automated inference
method; the stochastic gradient-based optimization algorithms that underpin
BBVI generally require extensive problem-specific tuning \cite{Schmidt21}. Such sensitivity
to tuning reduces the usability, robustness, and reliability of BBVI~\cite{Yao18,Dhaka20,Huggins20}.

This issue of tuning is of course not unique to VI, but is a fundamental problem for
the majority of gradient-based stochastic optimization algorithms~\cite{Schmidt21}.
The most common tuning task, and arguably most impactful on performance, is to set the \emph{step size} parameter
typical of most gradient-based stochastic optimization algorithms.
Too small and the algorithm makes little progress towards an optimum; too large and it can either diverge or fail to converge. 
There may not even be a single choice of step size that works well throughout the whole space.
This challenge has prompted significant attention from the stochastic optimization community over the last
decade, resulting in the development of a wide range of strategies for setting the step size
adaptively or removing it 
altogether \cite[e.g.,][]{Duchi11,Kingma15,Welandawe24,Reddi18,Vaswani20,De17,Orabona17,Defazio23,Ivgi23,Khaled23,Orabona21,Baydin18,Chen23,Cutkosky23,Galli23,Vaswani19}.
The goal of this work is to characterize the current landscape of these methods in the context of BBVI, and to determine
which, if any, can be used in BBVI to produce a 
truly automated and tuning-free methodology.\footnote{Concurrent work
\citep{Andersen26} provides an empirical study of default tuning settings
present in VI software packages (PyMC, NumPyro and TensorFlow Probability). The
study examines existing default choices of initialization, gradient estimate
batch size, and transformations for conjugate Bayesian models. The
surveyed packages use Adam and windowed Adagrad for optimization with a default
step size of $\approx 10^{-3}$ (see \citep[Table 4]{Andersen26}). By contrast,
we do not examine particular implementations of VI in software; rather we
investigate the performance of a broad collection of step-size adaptive
optimization algorithms in the context of variational inference. Many of these
algorithms have been developed recently and have not yet appeared in
open-source variational inference software packages. We develop appropriate
default settings for each that maximize performance across a wide range of
examples, and compare the tuned methods in the interest of identifying the best
candidates for future work on VI. Our benchmark set deliberately includes
difficult realistic applications which are not amenable to a theoretical
analysis.}

Towards this goal, we perform a large-scale empirical evaluation of 56 stochastic
gradient-based optimization algorithms applied to 1092 Bayesian inference
optimization problems, involving over 550,000 individual optimization runs
and roughly 15 core-years of compute. The algorithms we evaluate are chosen
to represent a wide spectrum of recent approaches to adaptive optimization.
The posterior targets are taken from a version of \texttt{posteriordb} \cite{posteriordb}
augmented to include noisy evaluation of log posterior densities and gradients due to data subsampling.
We consider four inference problems of increasing difficulty: the \emph{maximum a posteriori} (MAP) problem,
Gaussian VI with a diagonal covariance, Gaussian VI with a dense covariance, 
and VI with a neural-network based variational autoencoder family \cite{Kingma14-vae,Rezende14}.  
We first evaluate each algorithm with a range of step sizes to find a default value for each,
resulting in a set of tuning-free default optimizers.
We then compare the resulting default methods across our benchmark suite, including some challenging held-out benchmark problems, 
to characterize their general effectiveness.
We consider a wide range of performance metrics, including integrated and final
values of the ELBO and squared gradient norm, as well as rankings of each. Note that the
comparisons in this work are solely focused on the performance of 
optimization algorithms, as opposed to the quality of any variational family or
MAP approximation. In addition to evaluating existing optimization algorithms, our paper also
introduces two novel methods: a streaming variant of RABVI that overcomes its
large memory requirements \cite{Welandawe24} and sample-average approximation
(SAA) \cite{Kim15, Giordano24, Burroni24} with a batch-size increase rule
designed to avoid overoptimizing on small batches.

Our experiments reveal no clear single best default algorithm across all problems.
The top performer is perhaps Adam \cite{Kingma15} with step size $10^{-4}$,
but this algorithm ranks first among competitors on only $\sim 7\%$ of the problems, 
and is in the top 10 on only $\sim35\%$ of the problems.
The majority of algorithms also fail on more than $\sim20\%$ of problems of at least one objective type.
Surprisingly, however, we identify an ensemble of just 5 default methods that
rarely fails and consistently provides near-optimal performance among the set of 56 algorithms.
Specifically, the ensemble is: Adam with step
size $10^{-3}$ and $10^{-4}$, distance-over-weighted-gradient (DoWG) \cite{Khaled23} with step
size 1, Lion \cite{Chen23} with step size $10^{-5}$, and our novel variant of 
SAA-LBFGS with initial step size $10^{-8}$. 
This ensemble can be used as a solid baseline panel of 5 methods for follow-up
work on tuning-free optimization in VI, or can be used in practice as a single
ensemble optimizer in frameworks where expert tuning is not feasible.

\section{Benchmark Problems}\label{sec:problems}
Our test suite consists of a total of 1092 benchmark inference problems that span a range of variational families,
objective functions, condition numbers, dimensions, and gradient estimators typical of optimization-based inference in Bayesian statistics.
In this section, we detail how this collection of problems is constructed.

The test suite is built starting from \texttt{posteriordb v0.6.0} \cite{posteriordb}, a collection of
147 Bayesian posterior distributions arising from a wide range of applications. Each of these models is coded 
using the Stan probabilistic programming language \cite{stan}, which enables the pointwise evaluation of the log probability density
function $\log p(\cdot)$ and its gradient $\grad \log p(\cdot)$. Our tests involve all of these posterior distributions except one (\texttt{mnist-nn\_rbm1bJ100}), which 
was removed due to computational hardware limitations. 
To enable data subsampling during optimization---a common practice in stochastic variational inference introduced since~\cite{Titsias14}---we
have manually edited the code for 127 of the 146 posterior distributions to include an 
input for a term index $\log p(\cdot, i)$ such that 
$\log p(\cdot) = \sum_{i=1}^N \log p(\cdot, i)$, where $N$ is the number of log-probability density terms for that posterior (and likewise for the gradient; see 
\cref{lst:subsampledstan} in \cref{sec:data-subsampled-pdb}). This set of indexed Stan models is available at \url{https://github.com/trevorcampbell/subsampledposteriordb}.

\cref{fig:problem_scatter} shows a qualitative plot displaying three important characteristics of each posterior distribution:
the dimension $d$ of the posterior, the condition number $\kappa$, and the relative amount of data subsampling noise $\rho$.
More precisely, the condition number and relative noise were computed via
\[
\kappa = \frac{\lambda_{\text{max}}\lt(\Cov(X)\rt)}{\lambda_{\text{min}}\lt(\Cov(X)\rt)}, \quad
\rho = \sqrt{\frac{\Var(N\log p(X,I))}{\Var(\log p(X))}}, \quad
X \dist p, \quad
I \dist \Unif\{1,\dots, N\},
\] 
where all expectations were estimated using Markov chain Monte Carlo~\cite{Robert04} and $\lambda_{\text{max}},\lambda_{\text{min}}$ refer to the maximum 
and minimum eigenvalues of a matrix, respectively. Note that for those 19 posterior distributions where data subsampling
was not implemented, $\rho=1$.
\cref{fig:problem_scatter} demonstrates that the test suite contains problems exhibiting a wide range of
dimension and condition number, and that the subsampling noise is of 
a large enough magnitude to have a meaningful effect in many problems.

\bfig
\bsubfig{0.5\textwidth}
\includegraphics[width=\columnwidth]{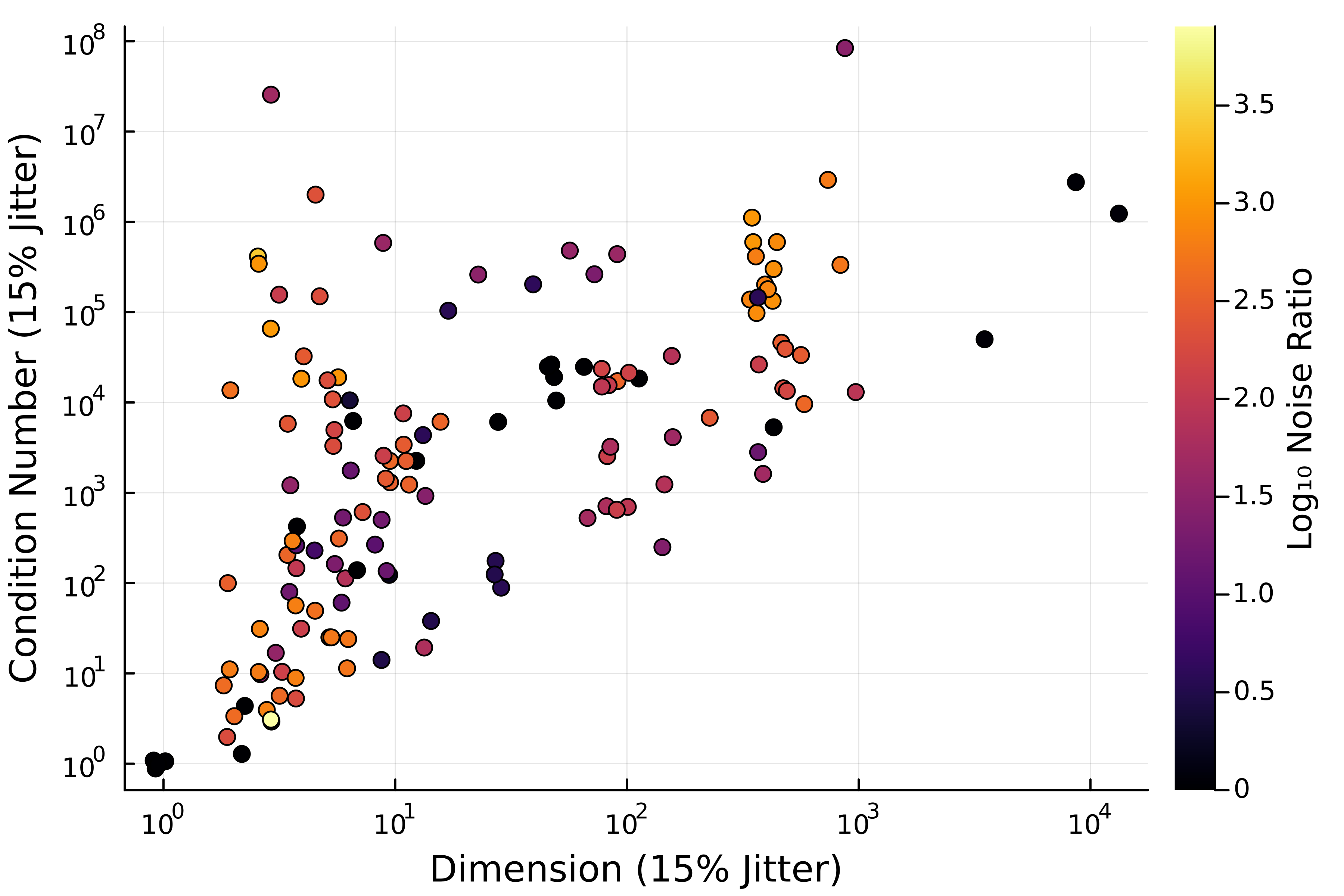}
\caption{}\label{fig:problem_scatter}
\esubfig
\bsubfig{0.5\textwidth}
\includegraphics[width=\columnwidth]{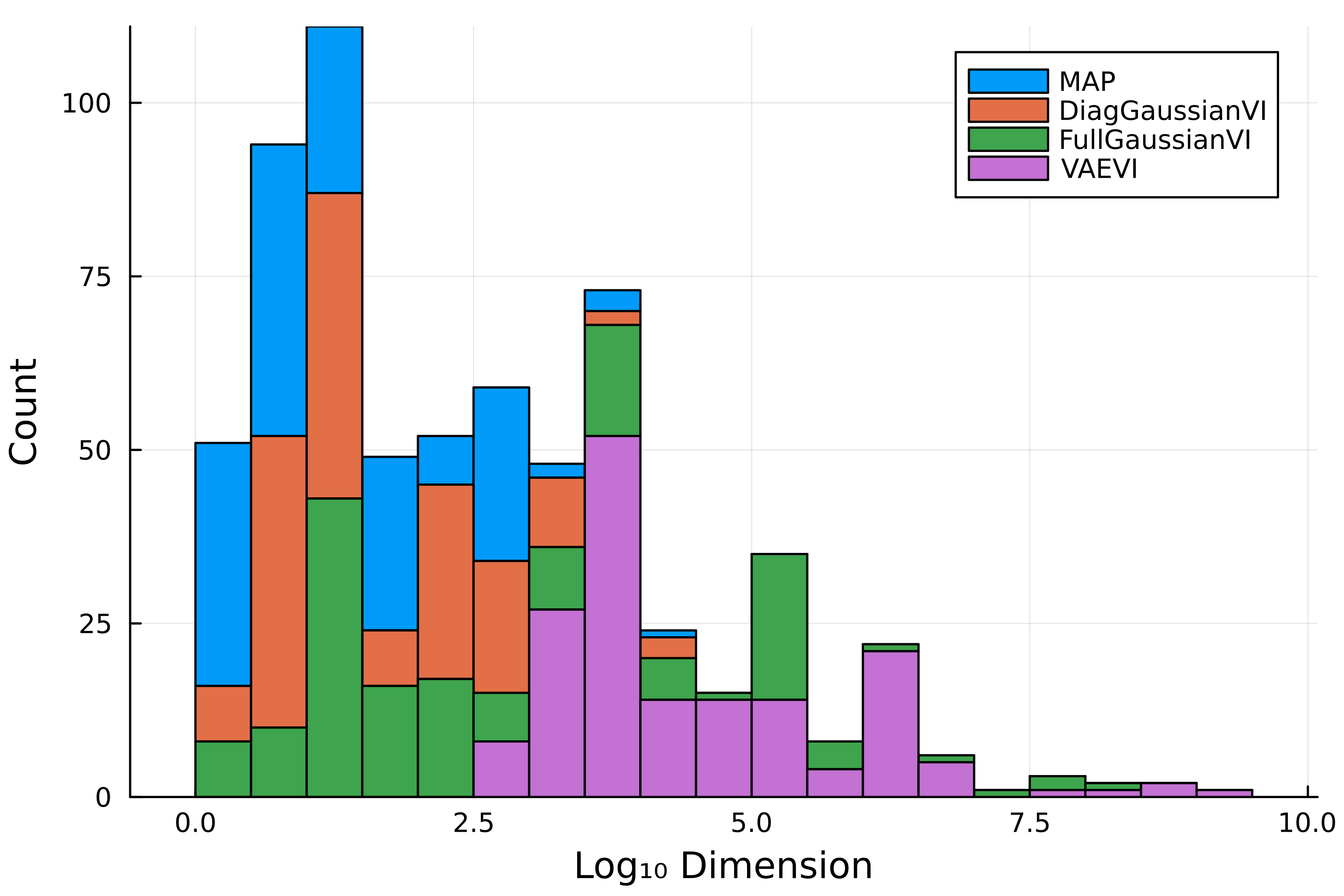}
\caption{}\label{fig:problem_histogram}
\esubfig
\caption{Qualitative characteristics of the problems in the benchmarking suite.
\cref{fig:problem_scatter}: Scatter of the dimension and covariance condition number of each of the 146 posterior targets, coloured by the relative noise standard deviation for data subsampling
versus no data subsampling. Dimension and condition number are jittered by 15\% to avoid overplotting.
\cref{fig:problem_histogram}: Stacked histogram of the dimension of the 1092 optimization problems, grouped by problem type.
}\label{fig:problem_properties}
\efig

The test suite includes 4 common objective functions for each of the 146
posterior distributions: maximum a posteriori (MAP) optimization, and
Kullback-Leibler divergence minimization with a diagonal Gaussian variational
family (DiagGaussianVI) \cite{Peterson89,Hinton93}, a dense covariance Gaussian variational family
(FullGaussianVI) \cite{Titsias14,Graves11,Kucukelbir17,Rezende14}, and a
neural-network-based variational autoencoder family (VAEVI) \cite{Kingma14-vae,Rezende14,Dayan95}. 
Furthermore, for those 127 distributions where we implemented data subsampling, we include the same
4 optimization problems again except where the gradient estimator includes subsampling noise.
This combination results in a total of $(127+146)\times 4 = 1092$ optimization problems.
\cref{fig:problem_histogram} shows the resulting stacked histogram of optimization problem dimensions,
demonstrating a wide range of nontrivial problems. We provide details on the objective
functions and gradient estimators below; recall that $d$ denotes the dimension of the posterior target,
and $N$ denotes the total number of data points in the subsampling cases.

\paragraph{MAP} Maximum a posteriori inference involves solving the $d$-dimensional optimization 
\[
\text{minimize}_{x\in\reals^d} &-\log p(x).
\]
For full-data problems, we use the exact gradient
\[
\shg(x) &= -\grad \log p(x),
\]
whereas for data-subsampled problems, we use the stochastic gradient estimate
\[
\shg(x, I) &= -N\grad\log p(x, I), \quad I\dist\Unif\{1,\dots,N\}.
\]
\paragraph{DiagGaussianVI} Variational inference with the diagonal Gaussian family involves solving the $2d$-dimensional optimization problem
\[
&\text{minimize}_{\mu, \sigma \in \reals^d} \KL\lt( \Norm\lt(\mu, \diag(\sigma_1^2,\dots,\sigma_d^2)\rt)||p\rt)\\
 \Leftrightarrow \qquad &\text{minimize}_{\mu, \sigma \in \reals^d} \;\Big\{\; C + \E\lt[-\log p(\mu + \sigma \circ Z)\rt] - \sum_{j=1}^d \log|\sigma_j| \;\Big\}
\]
where $Z\dist\Norm(0,I)$, $\circ$ denotes element-wise multiplication, and $C$ is a constant that does not depend on the optimization parameters.
For the full-data version of this problem, we use the ``closed-form entropy'' variant \cite{Titsias14} of the reparametrization gradient~\cite{Ho83,Rubinstein92,Kingma14-vae,Rezende14,Titsias14} estimator
\[
\shg_\mu(\mu,\sigma,Z) &= -\grad \log p(\mu+\sigma \circ Z)\\
\shg_\sigma(\mu,\sigma,Z) &= -\grad \log p(\mu+\sigma\circ Z)\circ Z - \frac{1}{\sigma},
\]
where $1/\sigma$ denotes the vector with entries $1/\sigma_j$.
For the data-subsampled version of this problem $\grad\log p(\mu+\sigma \circ Z)$ is replaced by $N\grad \log p(\mu+\sigma\circ Z, I)$ 
where $I\dist\Unif\{1,\dots, N\}$.

\paragraph{FullGaussianVI} Variational inference with the Gaussian family with a covariance with full-rank factors (``full-rank'' Gaussian) \cite{Titsias14,Graves11,Kucukelbir17,Rezende14} involves solving the $(d^2/2 + 3d/2)$-dimensional optimization problem
\[
&\text{minimize}_{\mu\in\reals^d, L\in\scL_d} \KL\lt(\Norm(\mu, LL^T)||p\rt)\\
\Leftrightarrow \qquad &\text{minimize}_{\mu\in\reals^d, L\in\scL_d} \Big\{\; C +  \E\lt[-\log p(\mu+ LZ)\rt] - \sum_{j=1}^d \log|L_{jj}| \;\Big\} \; ,
\]
where $Z\dist\Norm(0,I)$, $\scL_d$ is the set of $d\times d$ lower-triangular matrices, and $C$ is a constant that does not depend on the optimization parameters.
For the full-data version of this problem, we use the reparametrization gradient estimate
\[
\shg_\mu(\mu,\sigma,Z) &= -\grad \log p(\mu+LZ)\\
\shg_L(\mu,\sigma,Z) &= -\grad \log p(\mu+LZ) Z^T - (\diag L)^{-1}.
\]
Once again, for the data-subsampled version of the problem, $\grad \log p(\mu+LZ)$ is replaced by $N\grad \log p(\mu+LZ, I)$, where $I\dist \Unif\{1,\dots, N\}$.

\paragraph{VAEVI} Variational inference with a neural-network-based
autoencoder family involves solving the optimization problem
\[
\text{minimize}_{\lambda} \; &\KL\lt(q_\lambda || p_\lambda\rt),
\]
with augmented target $p_\lambda$ 
and variational approximation $q_\lambda$ on $\reals^d\times\reals^d$ given by
\begingroup
\interdisplaylinepenalty=10000
\[
\text{Augmented}&\text{ Variational Family $q_\lambda$}& \text{Augmented}&\text{ Target $p_\lambda$}\\
Z &\dist \Norm(0, I) &  X &\dist p\\
X \mid Z &\dist \Norm(\mu_\lambda(Z), \Sigma_{\lambda}(Z)) &  Z \mid X &\dist \Norm(m_\lambda(X), S_\lambda(X)) \label{eq:augvaefam}
\]
\endgroup
where $\mu_\lambda,\Sigma_\lambda,m_\lambda,S_\lambda$ are neural networks with tunable parameters $\lambda$.
In this work, $\Sigma_\lambda$ and $S_\lambda$ are diagonal matrices with entries $\sigma^2_{\lambda,j}$ and 
$s^2_{\lambda,j}$, respectively.
Note that the $X$-marginal of $p_\lambda$ is the original target distribution $p$.
We refer to $\mu_\lambda,\Sigma_\lambda$ as the \emph{decoder} (taking the ``latent'' $Z$ and returning the mean
and variance of the ``observed'' $X$) and $m_\lambda, S_\lambda$ as the \emph{encoder} (taking the ``observed'' $X$ and
returning the mean and variance of the ``latent'' $Z$), in analogy to the 
variational autoencoder \cite{Kingma14-vae,Rezende14,Dayan95}.
Note that in contrast to the traditional autoencoder setup, we can evaluate the density $p$ of $X$ up to normalization, but do not have draws of $X$, 
and so the role of the generative model and variational family are reversed as well as the direction of the KL divergence; otherwise, the problem setup is identical.
The encoder and decoder functions consist of 10-layer ResNets \cite{He16} with ReLU nonlinearities \cite{Nair10} and layer normalization \cite{Ba16},
resulting in an $r = (8d^2 + 844d)$-dimensional optimization problem; see \cref{alg:nn} in \cref{sec:neuralnets} for the detailed architecture.
Given the above choice of augmented target $p_\lambda$ and variational approximation $q_\lambda$,
we can expand the objective and rewrite the optimization problem as
\[
&\mathop{\text{minimize}}_{\lambda\in\reals^r} \E_{X,Z \sim q_\lambda}\bigg[\log\frac{\phi(Z;0, I) \phi(X; \mu_\lambda(Z), \Sigma_\lambda(Z))}{p(X)\phi(Z; m_\lambda(X), S_\lambda(X))}\bigg]\\
=&\mathop{\text{minimize}}_{\lambda\in\reals^r} \E\bigg[\!\!-\log p(x_\lambda(Z,\epsilon)) \!-\!\log \phi(Z | m_\lambda(x_\lambda(Z,\epsilon)), S_\lambda(x_\lambda(Z,\epsilon))) - \sum_{j=1}^d \log|\sigma_{\lambda,j}(Z)| \bigg],
\]
where $Z\dist \Norm(0,I)$, 
$\epsilon\dist \Norm(0,I)$ is used to reparametrize $X$ via 
$x_\lambda(Z,\epsilon) = \mu_\lambda(Z) + \Sigma_\lambda(Z)\epsilon$, and
$\phi(\cdot | \mu, \Sigma)$ is the density of a multivariate normal with mean $\mu$ and covariance $\Sigma$.
For the full-data problem, the gradient estimator is obtained by running automatic differentiation
on an estimate of the above objective based on a single draw of $Z,\epsilon$. 
For the data-subsampled problem, we replace $\log p(\cdot)$ with $N \log p(\cdot, I)$, $I\dist \Unif\{1,\dots, N\}$
and do the same with a single draw of $Z,\epsilon,I$.

\section{Algorithms}\label{sec:algorithms}

\btab
\caption{Table of Algorithms. Parameter settings for non-step-size parameters indicated below; most
are defaults taken from the respective original source papers, or default decay rates taken from Adam.
Note that HyperGradient, RABVI, SRABVI, and Mechanic are each 
applied to 8 SGD-likes (AdaGrad, Adam, AdamAvg, AMSGrad, DoG, DoWG, Lion, and SGD), so although there
are 28 rows in the table, there are 56 total unique algorithms in our test suite.}\label{tab:algs}
\btabr{c|c|c|c}
\textbf{Name} & \textbf{Type} & \textbf{Reference} & \textbf{Notes \& Parameters}\\
\hline
AdaGrad & SGD-Like & \cite[Eqn.~1]{Duchi11} & no parameters\\
Adam & SGD-Like & \cite[Alg.~1]{Kingma15} & defaults from Alg.~1\\
AdamAvg & SGD-Like & \cite[p.~13]{Welandawe24} & Adam defaults\\
AMSGrad & SGD-Like & \cite[Alg.~2]{Reddi18} & Adam defaults \\
AdaSLS & Line Search & \cite{Vaswani20} & (see text)\\
AdamSLS & Line Search & \cite{Vaswani20} & (see text)\\
BigBatch & Increasing Batch & \cite[Alg.~2]{De17} & Armijo $c=0.5$\\
COCOB & Other & \cite[Alg.~2]{Orabona17} & defaults from Alg.~2\\
DAdaptAdam & SGD-Like & \cite[Alg.~5]{Defazio23} & defaults from Alg.~5\\
DAdaptSGD & SGD-Like & \cite[Alg.~4]{Defazio23} & defaults from Alg.~4\\
DoG & SGD-Like & \cite{Ivgi23} & Initial movement $r_\epsilon=10^{-6}$\\
DoGMom & SGD-Like &  --- & DoG+momentum (decay $0.9$)\\
DoWG & SGD-Like & \cite[Alg.~1]{Khaled23} &Initial movement $r_\epsilon=10^{-6}$\\
DoWGMom & SGD-Like & --- & DoWG+momentum (decay $0.9$)\\
FTRL & Other & \cite[Alg.~1]{Orabona21} & defaults from Alg.~1\\
HyperGradient & Meta Tuner & \cite{Baydin18} & multiplicative scaling, $\beta = 10^{-4}$\\
Lion & SGD-Like & \cite{Chen23} & Adam defaults\\
Mechanic & Meta Tuner & \cite[Alg.~1]{Cutkosky23} & defaults from Alg.~1\\
Pflug & Meta Tuner & \cite[Alg.~4.1]{Pflug83} & no parameters\\
PoNoS & Line Search & \cite[Alg.~1]{Galli23} & (see text)\\
RABVI & Meta Tuner & \cite[Alg.~1]{Welandawe24} & defaults from VIABEL \cite{viabel}\\
SRABVI & Meta Tuner & --- & streaming RABVI (see text)\\
SAA & Increasing Batch & --- & (see text)\\
SAACG & Increasing Batch & --- & (see text)\\
SAALBFGS & Increasing Batch & --- & (see text)\\
SAANA & Increasing Batch & --- & (see text) \\
SGD & SGD-Like & \cite{Robbins51} & $1/\sqrt{t}$ schedule \\
SLS & Line Search & \cite[Alg.~1]{Vaswani19} & (see text)
\etabr
\etab

Our test suite includes of a total of 56 algorithms, listed in \cref{tab:algs}. In this section we describe the 
algorithms and any implementation changes that were necessary compared to the originals.
Many of these algorithms have multiple tuning parameters, all of which except the step size 
have been set to recommended defaults (see \cref{tab:algs}).
To aid discussion we split these algorithms up into 5 families of methods based on their coarse
algorithmic characteristics: 
stochastic-gradient-descent-like (SGD-like) methods,
stochastic line search algorithms,
increasing batch size methods,
meta-tuners,
and finally a few others 
that do not fit neatly in these categories. 
For brevity, we do not provide pseudocode for all of the methods in this paper; full 
Julia implementations of all methods can be found at \url{https://github.com/trevorcampbell/defaultvi}.

It is worth noting that some well-known optimization algorithms have been excluded from this work, 
as they are inapplicable to the full set of benchmark optimization problems and there is no straightforward way to adjust them.
Many algorithms \cite{Loizou21,Berrada20,Hazan22} require a known (lower bound on the) optimal objective function value, which is reasonable
for empirical risk minimization problems in machine learning but does not apply to variational inference
due to the unknown log normalizing constant. Others \cite{LeRoux12,Defazio14,Johnson13} require the gradient
noise to have finite support, which again rarely holds in variational inference (some exceptions are discussed in \cite{Kim24}) due to the expectation 
over the variational family in the objective.
Others still require a pre-specified, fixed number of iterations \cite{Vaswani22,Carmon22},
which we exclude because practitioners typically do not know how many iterations are needed in advance.
Optimization algorithms for variational inference in the black-box setting that are specific 
to certain variational families \cite{Domke19,Domke20,Domke23,Kim23,Modi23,Cai24,Cai24b,Lin19,Kumar25,Khan23,Lambert22,Diao23,Huszar17,Yu23,Tan25}
are excluded as they do not apply to the full range of problems we consider in this work.
Finally, there are certain stochastic optimization methods designed specifically for matrix-valued 
parameters \cite{Gupta18,Jordan24}, which we exclude to avoid requiring special 
knowledge about the structure of the optimization parameters.
Testing these in the context of variational inference 
is an interesting direction for follow-up work.

\paragraph{SGD-Like (12 Algorithms)}
SGD-like methods generally involve, at each iteration, drawing an unbiased
gradient estimate, making a simple online update to some statistics, and taking
a step using the gradient estimate and those statistics weighted by a step size
parameter.  These methods are typically quite straightforward to implement, and
have a low per-iteration cost. 
In this work, we include 12 SGD-like algorithms: AdaGrad
\cite{Duchi11}, Adam \cite{Kingma15}, AMSGrad \cite{Reddi18}, AdamAvg
\cite{Welandawe24}, Lion \cite{Chen23}, distance-over-gradient (DoG)
\cite{Ivgi23}, distance-over-weighted-gradient (DoWG) \cite{Khaled23},
d-adaptation versions of Adam and SGD \cite{Defazio23}, SGD with a $1/\sqrt{t}$ schedule, as well as the two
distance-over-gradient methods modified to include momentum (DoGMom, DoWGMom).
All of these methods are implemented as presented in their original source
material and do not require modification to suit the present context.
Non-step-size parameters are set to their defaults provided in the original source papers,
and Adam defaults are used for momentum/moment estimate decay rates where applicable; see \cref{tab:algs} for details.

\paragraph{Stochastic Line Search (4 Algorithms)}
Stochastic line search methods set their step size at each iteration using a
line search performed on the noisy objective estimate for that specific
iteration.  In this work we include 4 stochastic line search algorithms:
stochastic line search (SLS) \cite{Vaswani19}, AdaGrad with SLS step size
selection (AdaSLS) \cite{Vaswani20}, Adam with SLS step size selection
(AdamSLS) \cite{Vaswani20}, and Polyak non-monotone line search (PoNoS)
\cite{Galli23}.  Vanilla SLS is known to converge under mild conditions for
objectives exhibiting \emph{interpolation} \cite{Ma18,Vaswani19},
\textit{i.e.}, gradient variance that decays to 0 as the iterates approach the
optimum.  While this is unlikely to hold exactly in most variational problems,
it may be a reasonable approximation, especially in the transient phase of
optimization.  Also, certain types of gradient estimators---typically employing
some sort of control variate---can induce interpolation via variance reduction
\cite{Kim24}.  All line search methods are implemented using an Armijo line
search with coefficient 1/2, step size reduction factor 2, and incremental step
size reset factor 2 after each iteration.  PoNoS is implemented with $\xi = 1$
(in Eqn.~2), and the same simple incremental step size reset instead of the
Polyak step size reset described in the original paper.  We are unable to use
the Polyak step size reset, as it requires knowledge of the optimal objective
value for each minibatch function estimate.

Since all methods tend to forget their initial step size very quickly (within a
few stochastic iterations), we initialize the step size for these methods to
$10^{-6}$ and extend each method to include an ``outer step size'' that remains
constant over all iterations.  In other words, the update for these algorithms
is generalized to $x_{t+1}\gets x_t + \gamma \cdot \gamma_t \cdot g_t$, where
$\gamma_t$ is the original SLS-like step size tuned in each iteration using
line search, $g_t$ is the unscaled step direction, and $\gamma$ is a fixed
outer step size. Each algorithm recovers its original implementation when
$\gamma=1$, but we find smaller values often improve performance and reduce
failure rates.  Tuning results for these algorithms refer to the outer step
size $\gamma$.

\paragraph{Increasing Batch Size (5 Algorithms)}
Increasing batch size methods grow the number of draws used to produce a gradient estimate at each iteration until
some quality threshold is met. 
In this work, we include BigBatch \cite{De17}---which takes draws at each iteration until it achieves a particular signal-to-noise ratio
for the gradient estimate---and 4 variants of the sample average approximation (SAA) method, which 
optimizes a single, fixed estimate of the true objective function, and optionally increases the quality of the estimate over time \cite{Kim15}.
The variants include standard gradient descent with Armijo line search (SAA), Nesterov acceleration (SAANA), conjugate gradient (SAACG),
and LBFGS (SAALBFGS).
All methods involve line searches with Armijo constant $c=0.5$, step size shrinkage factor $2$, and step size resets with increase factor $2$.

Two SAA schemes have appeared recently in the variational literature, one involving a fixed batch size \cite{Giordano24} and one that grows the batch
using a rule based on hypothesis testing \cite{Burroni24}.
In this work, we grow the batch size using the following simple scheme. We initialize at iteration $t=0$ with sample size $n_0=1$. At each iteration $t$, the descent vector $g_t$
for batch size $n_t$ is compared with the descent vector $g'_t$ for batch size $2n_t$.
If $\|g_t\| < 0.5\|g'_t\|$ (not enough progress) or $g_t^{\top} g'_t < 0$ (wrong direction),
we set $n_t \gets 2n_t$ and repeat this test until it passes, at which point we take a step.
Pseudocode for these methods is given in \cref{alg:saa} in \cref{sec:saa}, with full implementation details
available at \url{https://github.com/trevorcampbell/defaultvi}.

It is worth noting that the sample approximation of the KL objective is usually degenerate until the batch size $n_t$ is large enough,
and starting with $n_0=1$ seems ill-advised.
For example, the DiagGaussianVI objective with $n_0 = 1$ takes the form $-\log p(\mu + \sigma \circ Z_1) - \sum_j \log |\sigma_j|$;
by fixing $\mu_j = c - \sigma_j Z_{1j}$ for some constant $c$, and increasing $\sigma_j$, we can send the approximate objective to $-\infty$.
In practice, we find the proposed termination criteria above increases the sample size to a reasonable value quickly and reliably.
For purely empirical work, this is satisfactory. For a theoretical guarantee that each optimization problem is well-posed, one might include a regularization
on the optimization parameters with a weight that decays quickly in $n_t$. 

\paragraph{Meta Tuners (5 Meta Algorithms, 33 Algorithms)}
Meta tuners are algorithms that run an inner base SGD-like algorithm, 
monitor its iterates, and modify its step size as that inner algorithm runs.
In this work we include 5 meta tuners: HyperGradient \cite{Baydin18},
robust and automated black-box VI (RABVI) \cite{Welandawe24},
a novel streaming variant of RABVI (SRABVI),
Mechanic \cite{Cutkosky23},
and Pflug's adaptive step size decay algorithm \cite{Pflug83}.
We apply the first 4 meta tuners to 8 SGD-like algorithms (AdaGrad, Adam, AdamAvg, AMSGrad, DoG, DoWG, Lion, and SGD),
and Pflug just to SGD per its original paper, resulting in 33 total algorithms in this group.
Mechanic and Pflug are implemented as described in their original source papers with no modifications.
HyperGradient is implemented using the multiplicative (scale-free) update in Eqn.~8, with adaptation step size $\beta=10^{-4}$.
RABVI was implemented without its termination criteria, and with parameters set as in the VIABEL repository \cite{viabel}.
Since RABVI has $O(T)$ memory cost for $T$ iterations, we find it often triggers out-of-memory errors;
we therefore include a streaming variant of RABVI (SRABVI) that uses $O(\log T)$ memory. Pseudocode for SRABVI
is given in \cref{alg:srabvi} in \cref{sec:srabvi}, and full implementation details are available at \url{https://github.com/trevorcampbell/defaultvi}.

\paragraph{Other (2 Algorithms)} We included two additional algorithms from the online optimization community:
follow-the-regularized-leader (FTRL) with a linearithmic regularizer \cite{Orabona21} and a $1/\sqrt{t}$ step size schedule,
and the backprop variant of continuous coin betting (COCOB) \cite{Orabona17}. Both of these algorithms have been
implemented as described in their original papers with recommended defaults. Note that COCOB has no tunable
step size parameter.

\section{Methodology \& Results}\label{sec:methodology}

In this section we describe the tuning and evaluation methodology and results for the algorithms in \cref{sec:algorithms}
applied to the benchmark test problems in \cref{sec:problems}.
All code was written in Julia \cite{Bezanson17} and run using version \texttt{1.12.4+0.x64.linux.gnu} on
UBC Advanced Research Computing's Sockeye cluster.
Posterior log probability density and gradient evaluations were provided by Stan \cite{stan} via the BridgeStan interface \cite{Roualdes23} and
PosteriorDB \cite{posteriordb} via PosteriorDB.jl \cite{posteriordbjl}.
All automatic differentiation was performed using the \texttt{Enzyme.jl} \cite{enzyme} package.
All code and results are available at \url{https://github.com/trevorcampbell/defaultvi}. 

\subsection{Experimental Methodology}
We ran each of the 56 optimization algorithms on each of the 1092 optimization problems, for a total of 61,152 jobs.
Within each job, we ran the algorithm with 9 step sizes---$10^{-8}, 10^{-7}, \dots, 10^{-1}$, and $10^{0}$---resulting in 
a total of 550,368 optimization runs. Each run was given the same initialization 
and a random number generator with the same initial seed.
Each run required 5 minutes for optimization, 
5 minutes for objective function evaluation, and 20 seconds for initial calibration
to account for potential differences in performance between nodes on the cluster.
On the cluster, we allocated 4GB of memory and 2 hours of total compute 
time for each job, which is slightly more than necessary for all the runs within a job. 
For each job that exceeded its memory or time budget, we re-ran that job with 48GB of memory 
and 6 additional hours of compute time. Any remaining jobs that exceeded time or memory constraints 
were treated as failures. The total compute used by this procedure was approximately 15 core-years.

Since the set of algorithms in \cref{sec:algorithms} are generally quite varied in nature, we use computation time 
as a common domain axis for comparison. Typical alternatives do not apply; for example, the number of iterations
is not comparable across methods with different per-iteration behaviour, and the number of log density/gradient evaluations requires empirical scaling to balance the 
two and does not capture the time cost involved in running the method itself. 
All algorithms were coded in Julia within one common framework, so there are no 
systematic differences between the methods that could arise, \textit{e.g.}, due to using different codebases by different authors in different languages.
To calibrate across potentially different compute nodes, prior to each run we repeatedly evaluated the gradient
at the initial point over 20 seconds, counted the number of completed evaluations, and incorporated that calibration into our results.

\bfig
\bsubfig{0.5\textwidth}
\includegraphics[width=\columnwidth]{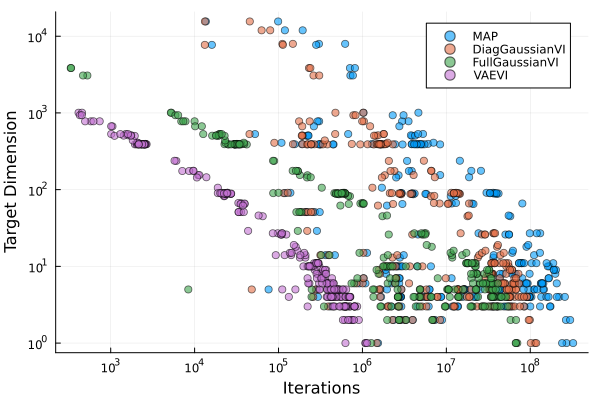}
\caption{}\label{fig:tuning_iteration_scatter}
\esubfig
\bsubfig{0.5\textwidth}
\includegraphics[width=\columnwidth]{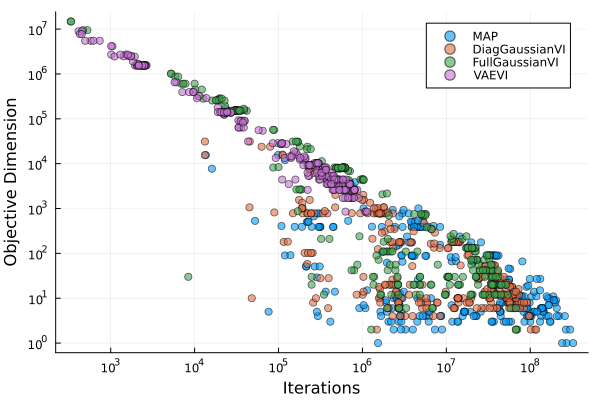}
\caption{}\label{fig:tuning_iteration_scatter_obj}
\esubfig
\caption{Scatter plots of the number of completed iterations versus target posterior dimension (\cref{fig:tuning_iteration_scatter}) 
and optimization parameter dimension (\cref{fig:tuning_iteration_scatter_obj}) during the tuning phase.
Iteration counts shown are the median across SGD-like methods, grouped by objective type. Iteration counts are equal to the 
number of gradient evaluations for SGD-like methods.
}\label{fig:tuning_iterations}
\efig

\cref{fig:tuning_iterations} displays the median number of completed iterations for each problem
across the SGD-like methods Adam, AdamAvg, DoG, DoWG, AMSGrad, AdaGrad, and SGD, showing its relationship
both with the target $\log p$ dimension and objective function dimension.
In the vast majority of problems, at least $10^4$ iterations were performed, which is a typical range for 
variational inference; some of the more difficult VAE and full Gaussian
VI experiments completed fewer. Note that for the SGD-like methods, the number of iterations and number of gradient evaluations is the same,
so these plots alternatively display the number of completed gradient evaluations.

During each run, we recorded two measures of performance at
logarithmically-spaced intervals: the estimated evidence lower bound (ELBO) and
squared gradient 2-norm, along with the estimated standard error in each. 
The number of draws used to estimate these performance measures was allocated
dynamically to each recorded time point to reduce the overlap between
confidence intervals of neighbouring estimates.
We record both the ELBO and squared gradient norm because neither is ideal in all situations.
The ELBO is the primary objective of interest in variational inference, but is complicated by the fact that 
it may exhibit local optima, and that the optimal value is an unknown real number.
Therefore, there is no straightforward way to produce relative ELBO performance comparisons across different problems.
On the other hand, the squared gradient norm is perhaps not of direct interest in variational inference, 
but is always nonnegative and has optimal value 0 making relative comparison straightforward.

We consider a run a \emph{hard failure} if one of four situations occurs: (1)
the run is incomplete due to an out-of-memory error, (2) the run is incomplete
due to an out-of-time error, (3) the Stan model encountered an exception while
evaluating $\log p$ or its gradient, or (4) the final ELBO objective function
is a numerical NaN/Inf or demonstrably larger than the initial
objective, in the sense that the final minus 1 standard error is greater than
the initial plus 1 standard error.  We consider a run a \emph{soft failure} if
one of two situations occurs: (1) it is a hard failure, or (2) the final
objective is not demonstrably smaller than the initial objective, in the sense
that the final plus 1 standard error is greater than the initial minus 1
standard error.

\subsection{Default Step Size Tuning}

To find a default parameter setting for each algorithm,
we use the following procedure. First, we remove all problems for which every algorithm and step size failed; this 
indicates a likely issue with the implementation of the model itself,\footnote{For example,
the Sucsceptible-Infected-Recovered (SIR) model posterior from PosteriorDB (\texttt{sir-sir}) involves an ODE integrator that can 
produce numerical near-0 negative values. These values are used as a Poisson mean parameter, which frequently causes Stan to raise an exception.}
or an extremely costly model for which no single run finished. 
Next, for each run, we rank the step sizes $10^{-8}, 10^{-7}, \dots, 10^{-1}, 10^{0}$
within that run according to their ELBO values at the 5-minute mark.
We focus on ranks as opposed to values to ensure comparability across different problems.
%We randomly shuffle the rankings among ties.

We then compute the average ranking for each step size across all runs, shown in \cref{fig:tuning_grid}. Since the ranks
are comparing the different step sizes for each algorithm individually,
\cref{fig:tuning_grid} should be interpreted row-wise; comparison should not 
be made across rows. Certain rows exhibit a large variations in colour---\textit{e.g.}, those for Adam, DoWG, DoG, and Lion---corresponding to
algorithms that prefer a particular step size as a default.  
Others exhibit very little variation---\textit{e.g.}, the SAA and Mechanic methods---corresponding to algorithms whose performance across the test suite
is insensitive to the step size parameter.
\cref{fig:failure_grid} visualizes the probability of (soft) failure for each algorithm
and step size across the test suite. Generally speaking, step sizes with a lower probability
of failure tend to correspond to those that are ranked more favourably.

\bfig
\includegraphics[width=\textwidth]{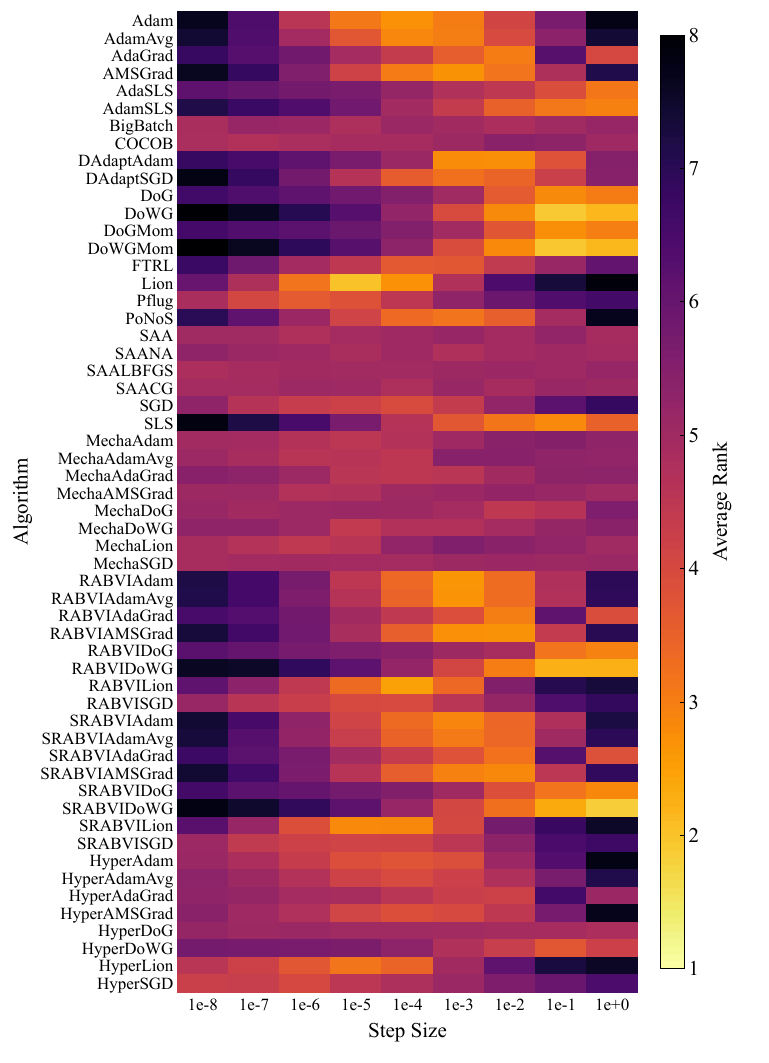}
\caption{Heatmap of the average rank (in terms of final ELBO objective) of each step size across the benchmark suite for each algorithm. Lower values (lighter colours) are better.
Each row of this figure pertains to one algorithm and is meant to be read by itself; do not compare across rows.
For algorithms with a wide range of colours in their row, the algorithm tends to prefer the step size with the lightest colour.
For algorithms with even colours across their row, the algorithm is insensitive to its step size setting and does not strongly prefer any particular value.
}\label{fig:tuning_grid}
\efig

\bfig
\includegraphics[scale=1.0]{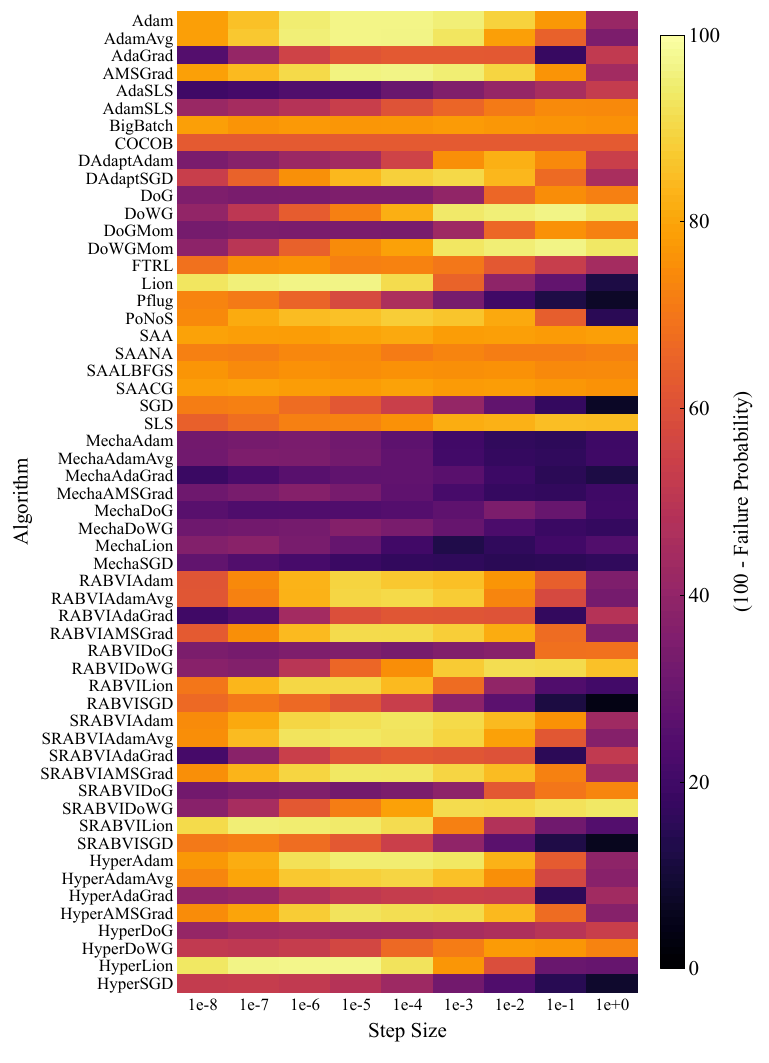}
\caption{
Heatmap of the probability that no soft failure occurs (including out-of-memory, out-of-time, Stan model exception, objective increase, and objective non-decrease failures)
for each step size across the benchmark suite for each algorithm. Higher values (lighter colours) are better.
Note that problems for which all algorithms and all step sizes failed have been removed from consideration.
Some algorithms have a range of step sizes with very light colours, indicating a region of reasonable values of the step size with a low failure probability.
Other algorithms have no light colours in their row, indicating a high rate of failures across the benchmark problem suite.
}\label{fig:failure_grid}
\efig

\btab
\caption{Recommended default tuning parameters for each algorithm in the test suite.
The numbers correspond to the $-\log_{10}$ step size with the minimum average
rank when considering (A) the final ELBO, (B) the integral of the ELBO, 
(C) the final square gradient norm, and (D) the integral of the square gradient norm.
Note that the integral objectives (B,D) tend to prefer quick decay of the objective via larger step sizes, while the final
objectives (A,C) tend to prefer convergence via smaller step sizes. As an overall recommendation we suggest using step size (A),
as it errs on the side of being slightly conservative while still providing good overall performance. Step size (A) is used
for all subsequent results pertaining to tuned algorithms. COCOB has no step size parameter, so has no default.}\label{tab:defaults}
{\scriptsize
\btabr{rllll}
 & \multicolumn{4}{l}{\textbf{$-\log_{10}($Step Size$)$}}\\
\textbf{Algorithm} &A&B&C&D\\
\hline
Adam&4&3&5&3\\
AdamAvg&4&3&4&4\\
AdaGrad&2&2&2&2\\
AMSGrad&3&3&4&3\\
AdaSLS&0&0&0&0\\
AdamSLS&0&0&0&0\\
BigBatch&5&3&5&3\\
COCOB&--&--&--&--\\ %&8&0&8&0\\
DAdaptAdam&2&2&2&2\\
DAdaptSGD&3&3&3&3\\
DoG&1&0&1&1\\
DoWG&1&1&1&1\\
DoGMom&1&0&1&0\\
DoWGMom&1&1&1&1\\
FTRL&4&4&4&4\\
Lion&5&5&5&5\\
Pflug&7&5&7&5\\
PoNoS&3&3&4&3\\
SAA& 0&3&3&3\\
SAANA&3&3&3&4\\
SAALBFGS&4&3&7&3\\
SAACG&5&3&3&3\\
SGD&5&5&5&5\\
SLS&2&2&3&2\\
MechaAdam&5&5&5&5\\
MechaAdamAvg&6&6&6&5\\
MechaAdaGrad&4&4&4&4\\
MechaAMSGrad&5&6&5&6\\
MechaDoG&1&2&1&1\\
MechaDoWG&4&4&4&4\\
MechaLion&6&6&6&6\\
MechaSGD&1&1&6&6\\
RABVIAdam&3&3&3&3\\
RABVIAdamAvg&3&3&3&3\\
RABVIAdaGrad&2&2&2&2\\
RABVIAMSGrad&3&3&3&3\\
RABVIDoG& 0&0&0&0\\
RABVIDoWG&1&1&1&1\\
RABVILion&4&4&4&4\\
RABVISGD&5&5&5&5\\
SRABVIAdam&3&3&3&3\\
SRABVIAdamAvg&4&0&1&6\\
SRABVIAdaGrad&2&2&2&2\\
SRABVIAMSGrad&3&3&3&3\\
SRABVIDoG&0&0&0&0\\
SRABVIDoWG&0&0&0&1\\
SRABVILion&5&4&5&4\\
SRABVISGD&5&5&5&5\\
HyperAdam&4&3&4&3\\
HyperAdamAvg&7&3&5&8\\
HyperAdaGrad&2&2&2&2\\
HyperAMSGrad&4&3&3&3\\
HyperDoG& 0 &0&0&0\\
HyperDoWG&1&1&1&1\\
HyperLion&5&4&5&4\\
HyperSGD&7&6&7&5\\
\etabr
}
\etab

Finally, we obtain the tuned default step sizes by minimizing the average rank in \cref{fig:tuning_grid}; 
the resulting $-\log_{10}($default step sizes$)$ are listed in \cref{tab:defaults} column A.
There are of course many alternative approaches that could have been used to select default step sizes.
Rather than the average 5-minute ELBO ranking, one might consider the median,
or the probability of rank 1, or the maximum of average ranks within each particular objective type.
Rather than the objective value at 5 minutes, one might consider the value at another time,
or the time-integral.
Rather than the ELBO, one might use the relative squared gradient norm, in which case
one might not use rankings but rather the values themselves.
Indeed we did examine many of these alternatives; we found the resulting
default parameter choices were generally insensitive to changing these criteria.
As an example, \cref{fig:gradrel_stepsizes} shows box plots of the relative squared 
gradient norm at 5 minutes, separated by objective type.
The trends here are essentially the same as in \cref{fig:tuning_grid}. For example,
Adam prefers step sizes $10^{-3}$--$10^{-4}$, DoWG prefers $10^{-1}$--$10^0$, and SAALBFGS
is insensitive to its step size.
Columns B--D in \cref{tab:defaults} show what default parameter setting we would obtain
if we minimized the average ranking of the time-integrated ELBO, the 5-minute squared gradient norm,
and the time-integrated squared gradient norm, all with similar results.
Certain algorithms do exhibit a large variation in default parameter setting---\textit{e.g.}, SAALBFGS has values 4,3,7,3, 
and MechaSGD has values 1,1,6,6---but this is a result of those algorithms' insensitivity to their step size parameter,
as opposed to sensitivity to the choice of tuning criterion, as demonstrated in the corresponding rows of \cref{fig:tuning_grid}.

\bfig
\bsubfig{\textwidth}
\includegraphics[scale=1]{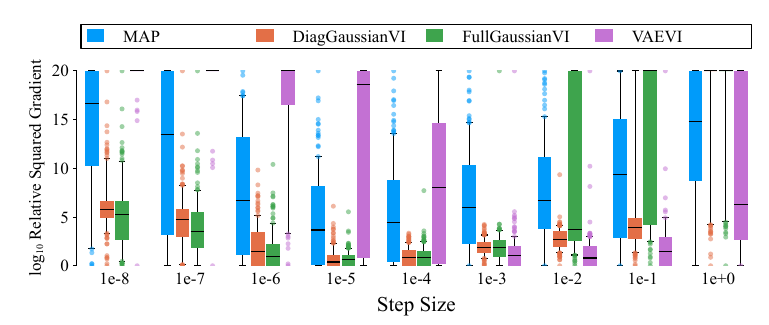}
\caption{Adam}\label{fig:gradrelsteps_adam}
\esubfig\\
\bsubfig{\textwidth}
\includegraphics[scale=1]{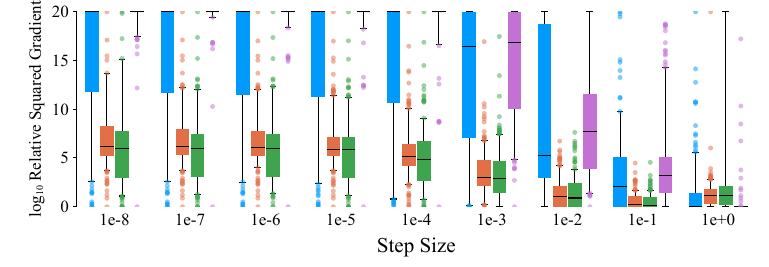}
\caption{DoWG}\label{fig:gradrelsteps_dowg}
\esubfig\\
\bsubfig{\textwidth}
\includegraphics[scale=1]{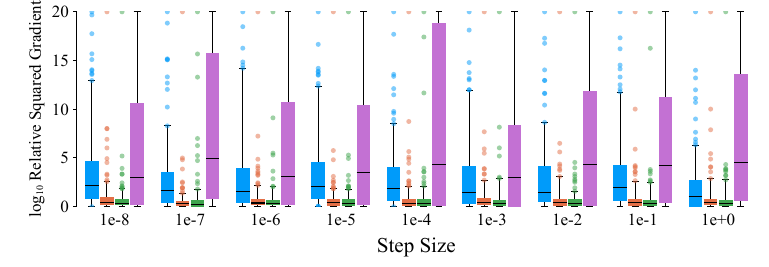}
\caption{SAALBFGS}\label{fig:gradrelsteps_saalbfgs}
\esubfig
\caption{
Box plots of squared gradient norm at the 5-minute mark relative to the minimum across step sizes,
grouped by objective type and
displayed for Adam, DoWG, and SAALBFGS. Lower values are better.
The whiskers demark the 80\% percentiles.
Trends for each of these algorithms are generally the same as the heatmap shown in \cref{fig:tuning_grid};
for example, Adam seems to prefer $10^{-4}$--$10^{-3}$, DoWG prefers $10^{-1}$--$10^{0}$, and SAALBFGS is insensitive
to its step size as the line search forgets its initial value quickly.
}\label{fig:gradrel_stepsizes}
\efig

\subsection{Evaluation of Default Algorithms and an Ensemble}

We now compare the default optimizers (with step sizes from \cref{tab:defaults} column A) 
across the problem suite. Note that the quality of the optimized variational families or MAP points
is not of interest in this work; what is of interest is the relative comparison
of the objective function traces produced by each method.
We focus again on comparing methods using the snapshot of their objective value
at the 5-minute mark; results seem to be qualitatively similar for other potential choices.
In this section we also consider the addition of one new method (Ensemble),
which consists of running five methods in parallel:
\[
\text{\textbf{Ensemble:} Adam}(10^{-3}),\text{ Adam}(10^{-4}),\text{ DoWG}(1),\text{ Lion}(10^{-5}),\text{SAALBFGS}(10^{-8}),
\]
and taking the best result from the five at each time
(where ``best result'' depends on which objective is being considered).
We selected these five methods based on a combination 
of performance across the different problem types and simplicity of implementation,
to ensure that others can easily incorporate the method as a benchmark in subsequent work.
More precisely, suppose we run each of the five algorithms independently, 
and at time $t$ the state and objective value estimate for each of the five algorithms
are given by $x^{(\text{A})}_t$ and $J^{(\text{A})}_t$, where 
$\text{A} \in\scA = \{\text{Adam3,Adam4,DoWG,Lion,SAALBFGS}\}$.
Then 
the objective function $J_t$ and state $x_t$ of the Ensemble method at time $t$ 
is given by
\[
J_t &= J^{(\text{Alg})}_t,
&
x_t &= x^{(\text{Alg})}_t,
&
\text{where}&
&
\text{Alg} &= \argmin_{\text{A}\in\scA} J^{(\text{A})}_t.
\]
The Ensemble method can be considered via two perspectives: as a panel of baseline
methods to include in future work on tuning-free methods, or as a single algorithm
that one should use in practice. 

The results of the comparison of tuned methods are displayed in 
\cref{fig:traceplotpanel,fig:default_vs_ensemble,fig:failure_bars,fig:rank_boxes,fig:rank_cdfs,fig:comparison_matrix_wins}.
We begin with an examination of performance of the tuned methods on a panel of individual problems.
To get an intuitive sense for what the optimization traces look like across all tuned algorithms,
\cref{fig:traceplotpanel} displays the performance of all tuned algorithms
on 8 optimization problems chosen uniformly randomly stratified by problem type
and data subsampling. 
For a quantitative comparison that incorporates the value of the objective function itself,
\cref{fig:default_vs_ensemble} displays box plots of the distribution of 5-minute squared gradient norm relative to the Ensemble
method across the whole problem suite. In this figure we report the squared gradient norm as opposed to the ELBO, which is 
not comparable across problems and potentially exhibits local optima.
\cref{fig:rank_boxes,fig:rank_cdfs,fig:comparison_matrix_wins} provide a quantative comparison of
the algorithms that ignores the objective values themselves and instead focuses on rankings, enabling a comparison based on ELBO.
One should be cautious about reading too far into the rank values, as ``near-ties'' can cause ranks to be quite noisy.
Finally,
\cref{fig:failure_bars} shows the probability of a soft or hard failure across the problem suite for each algorithm.

There are a few primary takeaways from these figures. First, the Ensemble method performs reliably well compared to the 56 methods across 
problem and noise types, despite itself only involving running 5 easily-implemented methods. Other methods generally perform poorly
in systematic ways; \textit{e.g.}, the increasing batch size methods generally do poorly on neural network problems, and many of the SGD-likes
do poorly on MAP problems. We view this as evidence that the panel of 5 methods comprising the Ensemble is a good set of baselines for future
work on tuning-free variational inference.
Second, there is no clear ``winner'' among the 56 original algorithms; even Adam, which is perhaps the best candidate, ranks 1st in terms of 
5-minute ELBO only 5-8\% of the time, and is in the top 10 only about 35\% of the time.
There seems to be room for additional work on tuning-free optimization for variational inference.
Finally, SAA methods perform surprisingly well on MAP, DiagGaussianVI, and FullGaussianVI problems, even in fairly high dimensions.

\bfig
\bsubfig{0.49\textwidth}
\includegraphics[scale=0.9]{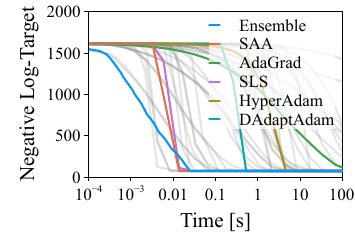}
\vspace{-1ex}
\caption{Bones (MAP)}
\vspace{-1ex}
\esubfig
\bsubfig{0.49\textwidth}
\includegraphics[scale=0.9]{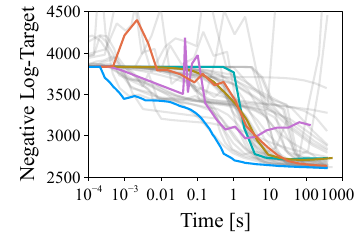}
\vspace{-1ex}
\caption{Science (MAP, subsamp.)}
\vspace{-1ex}
\esubfig
\\
\bsubfig{0.49\textwidth}
\includegraphics[scale=0.9]{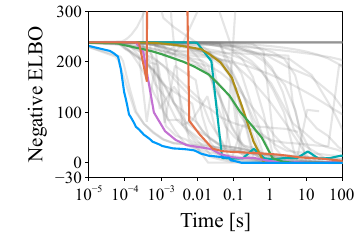}
\vspace{-1ex}
\caption{Pilots (DGVI)}
\vspace{-1ex}
\esubfig
\bsubfig{0.49\textwidth}
\includegraphics[scale=0.9]{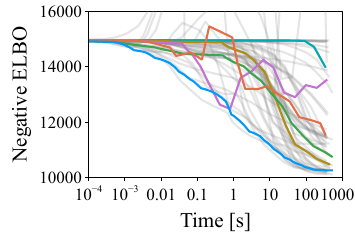}
\vspace{-1ex}
\caption{SAT (DGVI, subsamp.)}
\vspace{-1ex}
\esubfig
\\
\bsubfig{0.49\textwidth}
\includegraphics[scale=0.9]{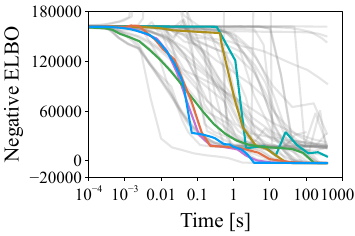}
\vspace{-1ex}
\caption{Diamonds (FGVI)}
\vspace{-1ex}
\esubfig
\bsubfig{0.49\textwidth}
\includegraphics[scale=0.9]{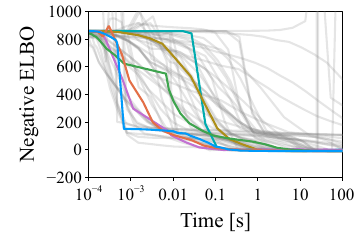}
\vspace{-1ex}
\caption{Mesquite (FGVI, subsamp.)}
\vspace{-1ex}
\esubfig
\\
\bsubfig{0.49\textwidth}
\includegraphics[scale=0.9]{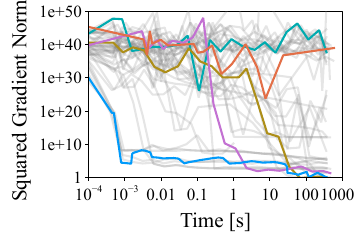}
\vspace{-1ex}
\caption{M0 (VAEVI)}
\vspace{-1ex}
\esubfig
\bsubfig{0.49\textwidth}
\includegraphics[scale=0.9]{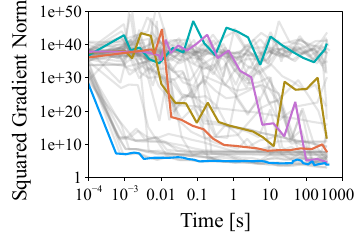}
\vspace{-1ex}
\caption{Dogs (VAEVI, subsamp.)}
\vspace{-1ex}
\esubfig
\caption{Example trace plots of performance on individual problems. For MAP problems the negative log target is plotted,
for DiagGaussanVI (DGVI) and FullGaussianVI (FGVI) problems the negative ELBO is plotted, and for VAEVI problems the squared gradient norm is plotted.
In all cases, lower values are better. Every tuned algorithm is displayed in each plot as a thin grey line, 
with Ensemble, SAA, AdaGrad, SLS, HyperAdam, and DAdaptAdam highlighted in 
colour to illustrate the performance of a variety of representative methods.}\label{fig:traceplotpanel}
\efig

\bfig
\includegraphics[scale=0.9]{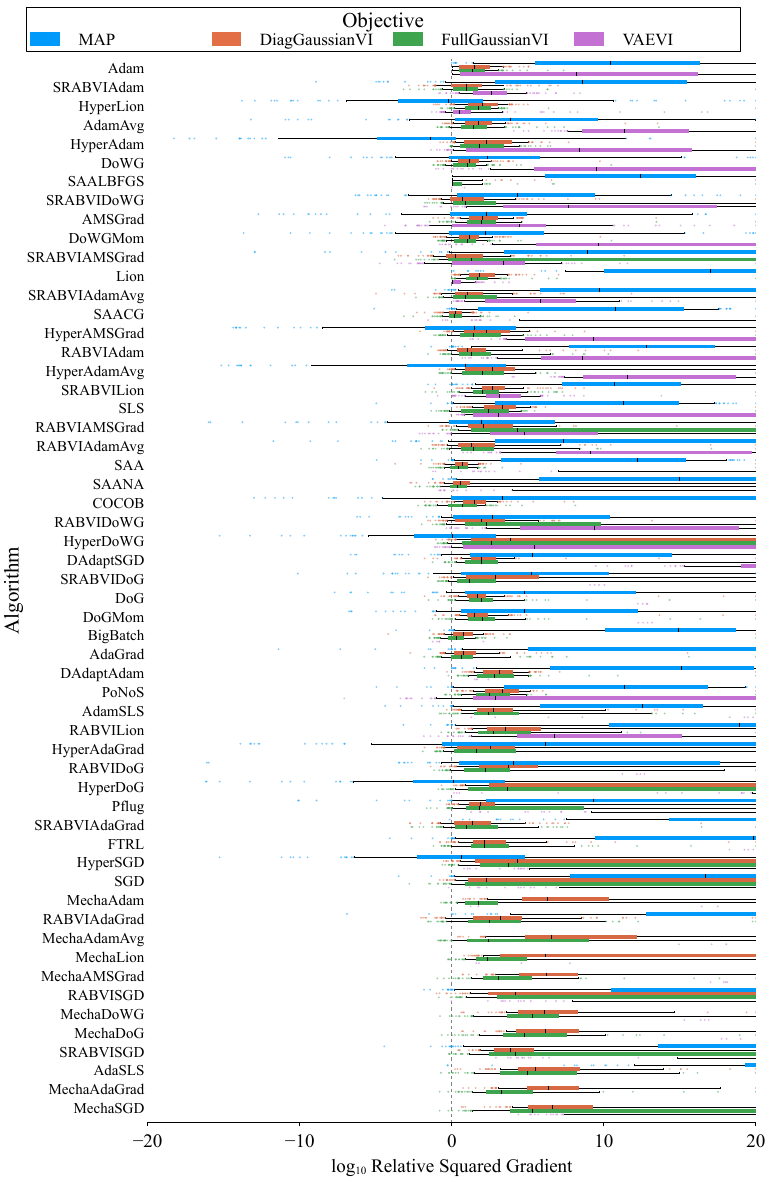}
\caption{Box plots of the 5-minute squared gradient norm of each tuned algorithm, relative to the Ensemble and split
by optimization problem type. The vertical grey dashed line at 0 indicates equal performance to the Ensemble method;
values to the right of the dashed line indicate worse
performance than Ensemble.
The whiskers demark the 80\% percentiles.
Note that values are thresholded to lie in the range $[-20,20]$ to
avoid plotting extreme values from failed runs. }\label{fig:default_vs_ensemble}
\efig

\bfig
\includegraphics[width=0.8\columnwidth]{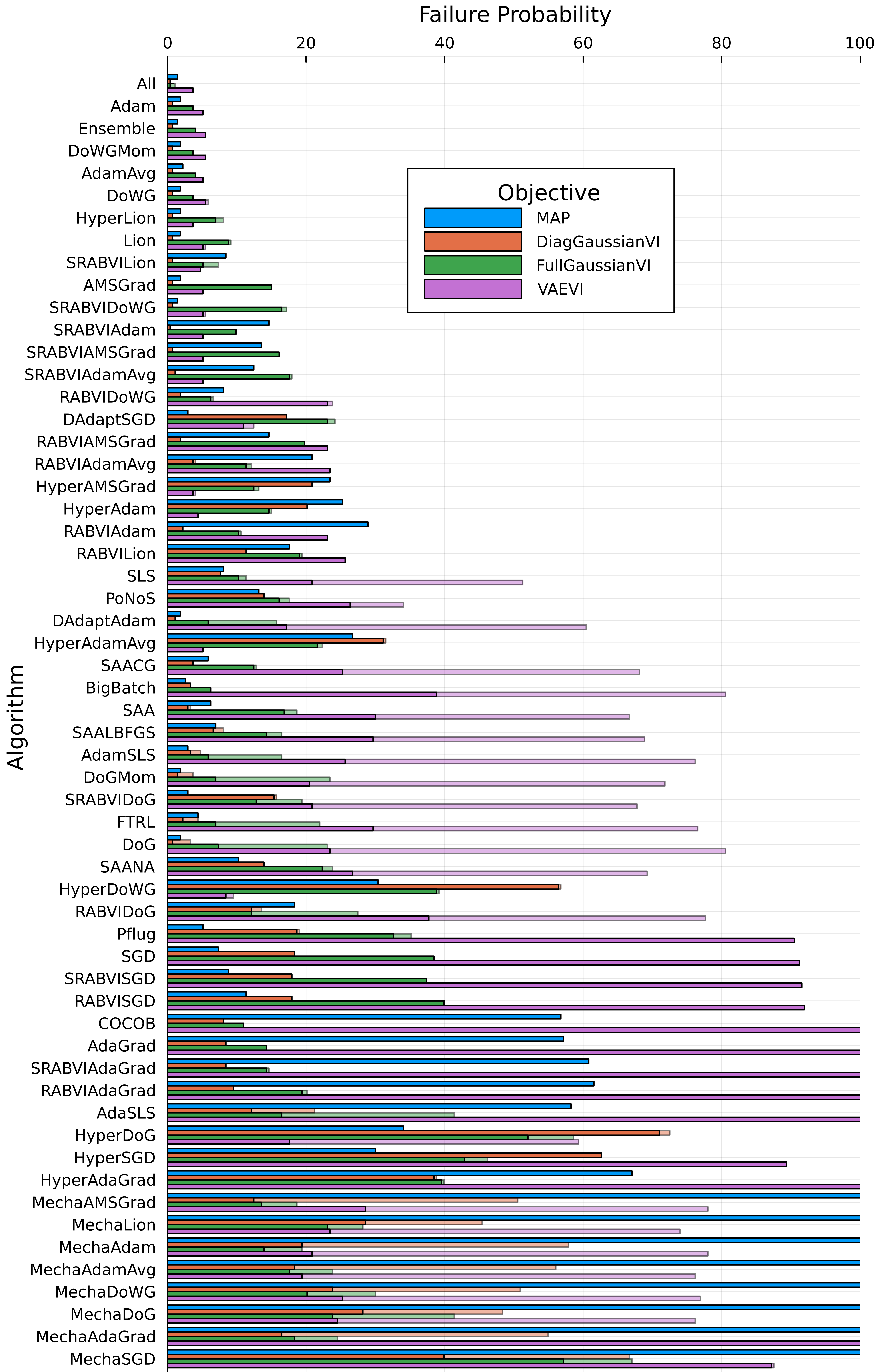}
\caption{Failure probability for each tuned algorithm grouped by objective function type, ordered by average failure probability. 
Dark bars display the probability of hard failure, and light bars display the probability of soft failure (there are always
at least as many soft failures as hard failures, because soft failures include hard failures).
The ``All'' group displays the amounts corresponding to optimization problems where every algorithm and step size failed, 
indicating a likely problem with the implementation of the model itself or a model so costly that no single run finished,
as opposed to failures caused by the optimization algorithms.}\label{fig:failure_bars}
\efig

\bfig
\includegraphics[scale=1]{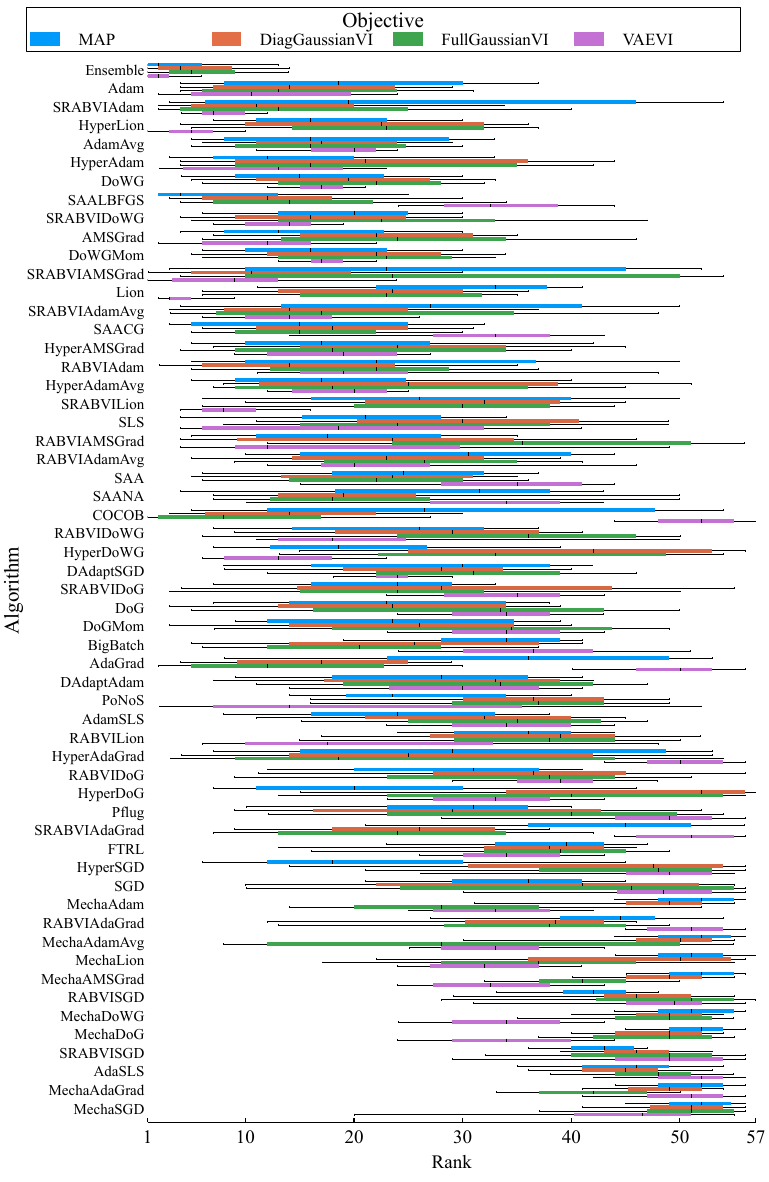}
\caption{Box plots of the 5-minute negative ELBO rank for each tuned algorithm, split by optimization problem type. The whiskers demark the 80\% percentiles.
Lower values (to the left) are better.}\label{fig:rank_boxes}
\efig

\bfig
\includegraphics[scale=1]{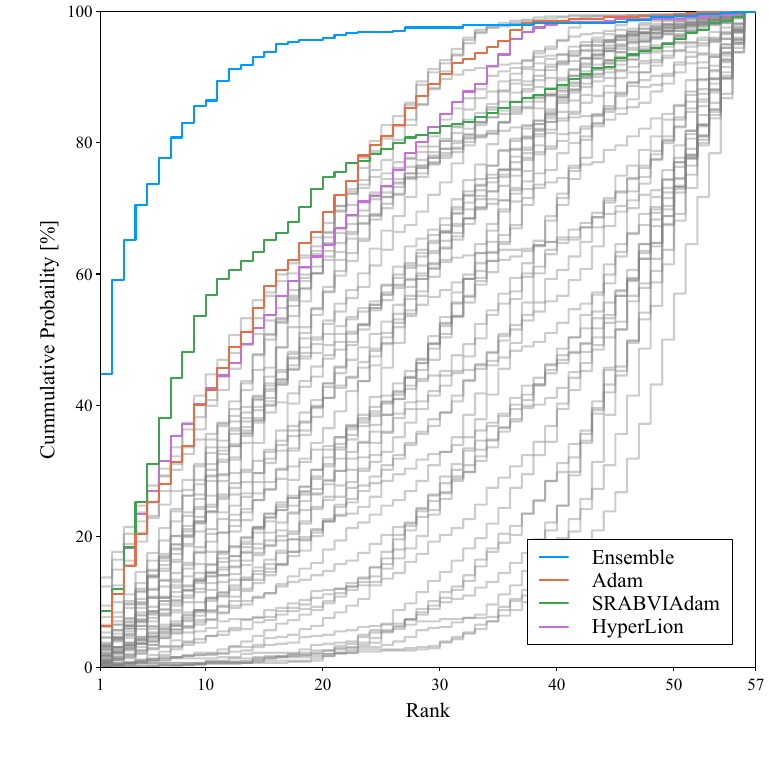}
\caption{Cumulative distribution functions of the 5-minute negative ELBO rank for all tuned algorithms.
Lines closer to the top left of the plot are better.
Most algorithms are shown as grey lines without a legend entry; a few representative algorithms are shown
in colour.}\label{fig:rank_cdfs}
\efig

\bfig
\includegraphics[width=\columnwidth]{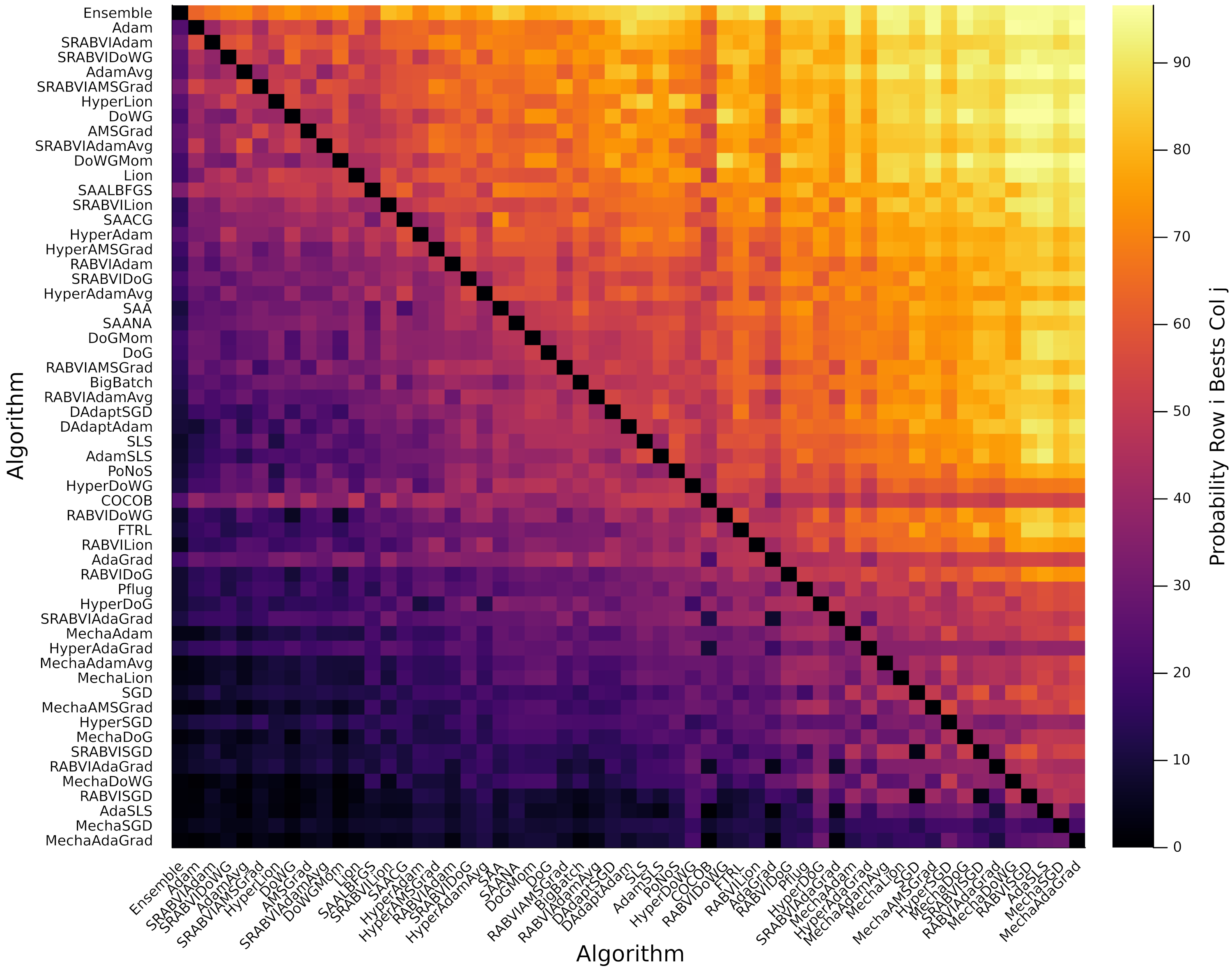}
\caption{Heatmap showing the probability that each tuned algorithm in row $i$ has a lower 5-minute negative ELBO than that of the algorithm in column $j$
across all optimization problems. Lighter colours indicate a higher probability that row $i$ performs better than row $j$.}\label{fig:comparison_matrix_wins}
\efig

\subsection{Evaluation on Challenge Problems}

Finally, we compare the default optimizers on a small number of challenge problems from beyond
\texttt{posteriordb}, as a form of ``held-out'' validation that the Ensemble remains a top performer on problems
not seen during bulk tuning:
\begin{itemize}
\item  {\bf NMF-Lung ($n=38$, $p=96$, target posterior dimension $d=3375$)} is a nonnegative matrix factorization model with a Poisson likelihood and sparsity-inducing
prior for the loadings \citep{Zito:2024}, fitted whole genome sequencing data for 38 lung adenocarcinoma tumor samples from the PCAWG project. 
We used single-base substitution counts and following the model hyperparameter choices from \citet{Xue:2024}. 
\item  {\bf NMF-Stomach ($n=75$, $p=96$, target posterior dimension $d=4300$)} is the same as the NMF-Lung model, except the data is for 75 stomach tumors. 
\item  {\bf NMF-Skin ($n=107$, $p=96$, target posterior dimension $d=5100$)}  is the same as the NMF-Lung model, except the data is for 107 melanoma tumors. 
\item {\bf SpLog-Leukemia ($n=72$, $p=7129$, target posterior dimension $d=14261$)} is a sparse logistic regression model
applied to a leukemia dataset. 
\item {\bf SIRNB ($n=14$, target posterior dimension $d=3$).} A susceptible-infectious-recovered (SIR) model with a negative binomial likelihood,
fitted to data from a 1978 influenza outbreak \cite{outbreak}.
The likelihood is parameterized by nonlinear ODE which needs to be solved numerically at each gradient evaluation.
\item {\bf 2CP ($n=1020$, target posterior dimension $d=111$).} A two-compartment population pharmacokinetic (PK) model.
The model describes the absorption and clearance of a drug compound in  20 patients via a linear ODE.
PK parameters are estimated for each patient and partially pooled via a hierarchical prior.
The model is fitted to simulated data following \citet{Margossian22}.
\end{itemize}

The results of this exercise are shown for a representative subset of variational family types and 
types of stochasticity in \cref{fig:heldtraceplotpanel}. The results mostly reflect those from the earlier
large-scale test; the Ensemble performs well across the board, and other methods perform poorly in various systematic
ways. For example, COCOB performs quite well on certain Gaussian VI problems
(\textit{e.g.}, the grey line to the left of the group in \cref{subfig:cocob1,subfig:cocob2}), but has a high chance of failure
on other kinds of problems (see \cref{fig:failure_bars}).

\bfig
\bsubfig{0.49\textwidth}
\includegraphics[scale=0.9]{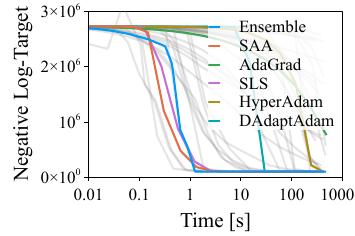}
\vspace{-1ex}
\caption{NMF-Stomach (MAP)}
\vspace{-1ex}
\esubfig
\bsubfig{0.49\textwidth}
\includegraphics[scale=0.9]{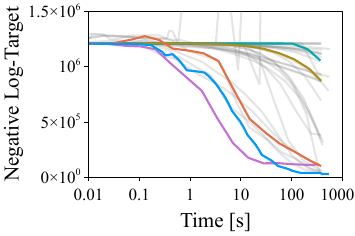}
\vspace{-1ex}
\caption{NMF-Skin (MAP, subsamp.)}
\vspace{-1ex}
\esubfig
\\
\bsubfig{0.49\textwidth}
\includegraphics[scale=0.9]{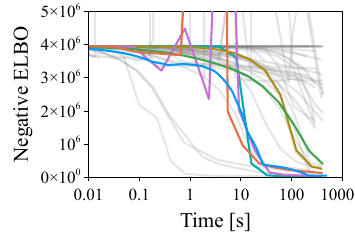}
\vspace{-1ex}
\caption{NMF-Lung (DGVI)}\label{subfig:cocob1}
\vspace{-1ex}
\esubfig
\bsubfig{0.49\textwidth}
\includegraphics[scale=0.9]{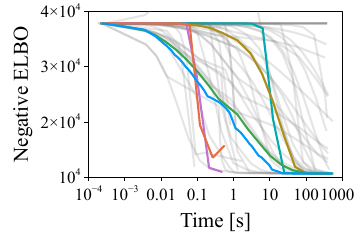}
\vspace{-1ex}
\caption{SpLog-Leukemia (DGVI, subsamp.)}
\vspace{-1ex}
\esubfig
\\
\bsubfig{0.49\textwidth}
\includegraphics[scale=0.9]{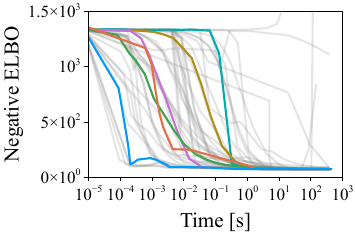}
\vspace{-1ex}
\caption{SIRNB (FGVI)}
\vspace{-1ex}
\esubfig
\bsubfig{0.49\textwidth}
\includegraphics[scale=0.9]{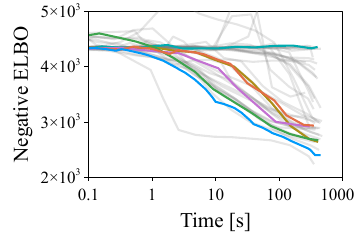}
\vspace{-1ex}
\caption{2CP (FGVI, subsamp.)}\label{subfig:cocob2}
\vspace{-1ex}
\esubfig
\\
\bsubfig{0.49\textwidth}
\includegraphics[scale=0.9]{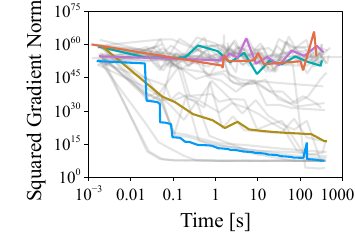}
\vspace{-1ex}
\caption{2CP (VAEVI)}
\vspace{-1ex}
\esubfig
\bsubfig{0.49\textwidth}
\includegraphics[scale=0.9]{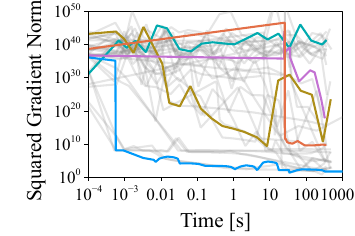}
\vspace{-1ex}
\caption{SIRNB (VAEVI, subsamp.)}
\vspace{-1ex}
\esubfig
\caption{Trace plots of performance on the held-out challenge problems. For MAP problems the negative log target is plotted,
for DiagGaussanVI (DGVI) and FullGaussianVI (FGVI) problems the negative ELBO is plotted, and for VAEVI problems the squared gradient norm is plotted.
In all cases, lower values are better. Every tuned algorithm is displayed in each plot as a thin grey line, 
with Ensemble, SAA, AdaGrad, SLS, HyperAdam, and DAdaptAdam highlighted in 
colour to illustrate the performance of a variety of representative methods.}\label{fig:heldtraceplotpanel}
\efig

\section{Conclusions}

This work presents a large-scale empirical test of 56 optimization algorithms on 1092 benchmark 
variational and maximum a posteriori Bayesian inference problems, with the goal of establishing good default tuning
parameters for each and subsequently establishing the state of the art in tuning-free optimization for Bayesian inference.
Our contributions include this set of reasonable default step size parameters for each optimization algorithm,
a simple ensemble method that reliably obtains near-optimal performance compared to the 56 algorithms in our test suite, 
a suite of 127 data-subsampled log posterior target densities, 
and a full release of all code and results from over 550,000 optimization runs for independent analysis. 
In an abstract perspective, our experiments are evaluating the performance of various stochastic gradient-based 
optimization algorithms in a statistical context. We expect our conclusions to
generalize reasonably to other BBVI schemes that utilize more elborate gradient
estimators~\cite{Roeder17,Mohamed20,Buchholz18,Fujisawa21}, alternative
variational objectives~\cite{Burda15,Domke18,Hernandezlobato16,Dieng17,Geffner21,Zhang21,Domke19-dividecouple,Salimans15,Li16},
and variational families~\cite{Rezende15,Domke18,Ong18,Ranganath16,Titsias19,Tan18}.

One key takeaway from this paper is that an ensemble of 5 methods consisting of
Adam with step size $10^{-3}$, Adam with step size $10^{-4}$,
DoWG with step size $1$, Lion with step size $10^{-5}$, and SAALBFGS with step size $10^{-8}$ generally outperforms the whole original 
set of 56 optimization algorithms. In future work, this ensemble could certainly be implemented as its own method given the availability of
sufficient hardware parallelism. However, we view the value of this method more as a reduction in the work required for comparisons in future work:
the ensemble is more likely to find fruitful use as a solid baseline panel of 5 methods against which one should 
compare other proposed tuning-free optimizers. Note that when conducting such comparisons, it is important to judge the performance 
of any proposed tuning-free method across as many problems as possible.

It is worth noting that the suggested default step sizes from this paper are not meant to be a panacea, and likely depend
on the particular set of problems and variational families selected for this work, however broad. That being said, these defaults should find value
in three scenarios: (1) as good default step sizes when expert tuning is expensive or not possible (\textit{e.g.}, in a practitioner-friendly probabilistic programming ecosystem),
(2) as step size values to use when comparing to genuinely tuning-free methods in follow-up research, and
(3) as good starting points for expert tuning on individual problems.

%%%%%%%%%%%%%%%%%%%%%%%%%%%%%%%%%%%%%%%%%%%%%%
%% Acknowledgements                         %%
%% should be provided in the                %%
%% Acknowledgements section.                %%
%%%%%%%%%%%%%%%%%%%%%%%%%%%%%%%%%%%%%%%%%%%%%%
\begin{acks}[Acknowledgments]
This research was supported in part through the computational resources and services provided by 
Advanced Research Computing at the University of British Columbia, notably the Sockeye compute cluster.
\end{acks}

%%%%%%%%%%%%%%%%%%%%%%%%%%%%%%%%%%%%%%%%%%%%%%
%% Funding information, if any,             %%
%% should be provided in the                %%
%% funding section.                         %%
%%%%%%%%%%%%%%%%%%%%%%%%%%%%%%%%%%%%%%%%%%%%%%
\begin{funding}
T.~Campbell and C.~Margossian were funded by National Sciences and Engineering Research Council of Canada (NSERC)
Discovery Grants. 
J.~H.~Huggins was partially supported by National Science Foundation CAREER award IIS-2340586 and Macrosystems Biology award DEB-2406258. 
\end{funding}

%%%%%%%%%%%%%%%%%%%%%%%%%%%%%%%%%%%%%%%%%%%%%%
%% Supplementary Material, including data   %%
%% sets and code, should be provided in     %%
%% {supplement} environment with title      %%
%% and short description. It cannot be      %%
%% available exclusively as external link.  %%
%% All Supplementary Material must be       %%
%% available to the reader on Project       %%
%% Euclid with the published article.       %%
%%%%%%%%%%%%%%%%%%%%%%%%%%%%%%%%%%%%%%%%%%%%%%
\begin{supplement}
\stitle{Appendices}
\sdescription{Descriptions of how Stan posteriors were modified to enable data subsampling,
detailed pseudocode for the neural networks used in the VAE experiments,
detailed pseudocode for the SAA methods,
and detailed pseudocode for the modified streaming RABVI algorithm.}
\end{supplement}
\begin{supplement}
\stitle{Subsampled PosteriorDB}
\sdescription{An online repository of Stan code for subsampled versions of posteriors from \texttt{posteriordb}. 
Available at \url{https://github.com/trevorcampbell/subsampledposteriordb}.}
\end{supplement}
\begin{supplement}
\stitle{Code and Results}
\sdescription{An online repository storing both the code to run the experiments in this paper as well
as an archive of all results for further independent examination.
Available at \url{https://github.com/trevorcampbell/defaultvi}.}
\end{supplement}

\bibliographystyle{ba}
\bibliography{sources}

@article{Andersen26,
  author = {Madelyn Andersen and Elizabeth Bersson and Tamara Broderick},
  year = {2026},
  title = {The curious case of the default settings: evaluating default performance of variational inference software},
  journal = {arXiv:2608.01403v1}}

@article{Dieuleveut20, 
year = {2020}, 
title = {bridging the gap between constant step size stochastic gradient descent and {{Markov}} chains}, 
author = {Dieuleveut, Aymeric and Durmus, Alain and Bach, Francis}, 
journal = {The Annals of Statistics}, 
pages = {1348 -- 1382}, 
number = {3}, 
volume = {48}, 
}

@article{Vehtari21,
  author = {Aki Vehtari and Andrew Gelman and Daniel Simpson and Bob Carpenter and Paul-Christian B\"urkner},
  title = {Rank-normalization, folding, and localization: An improved $\widehat{R}$ for assessing convergence of {MCMC} (with Discussion)},
  journal = {Bayesian Analysis},
  volume = {16},
  number = {2},
  pages = {667--718},
  year = {2021}}

@article{Welford62,
  author = {B. P. Welford},
  title = {Note on a method for calculating corrected sums of squares and products},
  journal = {Technometrics},
  volume = {4},
  number = {3},
  pages = {419--420},
  year = {1962}}

@article{Chan83,
  author = {Tony F. Chan and Gene H. Golub and Randall J. LeVeque},
  title = {Algorithms for computing the sample variance: {{Analysis}} and recommendations},
  journal = {The American Statistician},
  volume = {37},
  number = {3},
  pages = {242--247},
  year = {1983}}

@article{Flegal10,
  author = {James M. Flegal and Galin L. Jones},
  title = {Batch means and spectral variance estimators in {M}arkov chain {M}onte {C}arlo},
  journal = {The Annals of Statistics},
  volume = {38},
  number = {2},
  pages = {1034--1070},
  year = {2010}}

@article{Vats19,
  author = {Dootika Vats and James M. Flegal and Galin L. Jones},
  title = {Multivariate output analysis for {M}arkov chain {M}onte {C}arlo},
  journal = {Biometrika},
  volume = {106},
  number = {2},
  pages = {321--337},
  year = {2019}}

@article{Liu22,
  author = {Ying Liu and Dootika Vats and James M. Flegal},
  title = {Batch size selection for variance estimators in {MCMC}},
  journal = {Methodology and Computing in Applied Probability},
  volume = {24},
  number = {1},
  pages = {65--93},
  year = {2022}}

@article{Polyak92,
  author = {Boris T. Polyak and Anatoli B. Juditsky},
  title = {Acceleration of stochastic approximation by averaging},
  journal = {SIAM Journal on Control and Optimization},
  volume = {30},
  number = {4},
  pages = {838--855},
  year = {1992}}

@article{Xue:2024,
year = {2024}, 
title = {{Robust discovery of mutational signatures using power posteriors}}, 
author = {Xue, Catherine and Miller, Jeffrey W. and Carter, Scott L. and Huggins, Jonathan H.}, 
journal = {bioRxiv}, 
doi = {10.1101/2024.10.23.619958}, 
pages = {2024.10.23.619958}, 
}

@article{Zito:2024,
year = {2024}, 
title = {Compressive {{Bayesian}} non-negative matrix factorization for mutational signatures analysis}, 
author = {Zito, Alessandro and Miller, Jeffrey W}, 
journal = {arXiv:2404.10974}
}

@inproceedings{posteriordb,
  title = {{{posteriordb}}: {{Testing}}, benchmarking and developing {{Bayesian}} inference algorithms},
  shorttitle = {Posteriordb},
  booktitle = {Proceedings of {{The International Conference}} on {{Artificial Intelligence}} and {{Statistics}}},
  author = {Magnusson, M{\aa}ns and Torgander, Jakob and B{\"u}rkner, Paul-Christian and Zhang, Lu and Carpenter, Bob and Vehtari, Aki},
  year = 2025,
  series = {{{PMLR}}},
  volume = {258},
  pages = {1198--1206},
}

@misc{posteriordbjl,
	author = {Seth Axen},
	year = {2026},
	title = {PosteriorDB.jl: A {{Julia}} package to work with posteriordb},
	howpublished = {GitHub Repository: \url{https://github.com/sethaxen/PosteriorDB.jl}},
	note = {Version 0.6.0}}

@misc{stan,
	author = {{Stan Development Team}},
	year = {2026},
	title = {Stan Reference Manual, v2.38},
	howpublished = {\url{https://mc-stan.org}}}

@inproceedings{Loizou21,
  title = {Stochastic {{Polyak}} step-size for {{SGD}}: {{An}} adaptive learning rate for fast convergence},
  shorttitle = {Stochastic Polyak Step-Size for {{SGD}}},
  booktitle = {Proceedings of {{The International Conference}} on {{Artificial Intelligence}} and {{Statistics}}},
  author = {Loizou, Nicolas and Vaswani, Sharan and Laradji, Issam Hadj and {Lacoste-Julien}, Simon},
  year = 2021,
  series = {{{PMLR}}},
  pages = {1306--1314},
}

@inproceedings{Berrada20,
	author = {Leonard Berrada and Andrew Zisserman and {M.~Pawan} Kumar},
	title = {Training neural networks for and by interpolation},
	year = {2020},
	booktitle = {Proceedings of the International Conference on Machine Learning},
        series = {{PMLR}},
        volume = {119},
        pages = {799--809},
        }

@article{Hazan22,
	author = {Elad Hazan and Sham Kakade},
	title = {Revisiting the {{P}}olyak step size},
	year = {2019},
	journal = {arXiv:1905.00313}}

@inproceedings{Galli23,
 author = {Leonardo Galli and Holger Rauhut and Mark Schmidt},
 title = {Don't be so monotone: {{Relaxing}} stochastic line search in over-parametrized models},
 year = {2023},
 booktitle = {Advances in Neural Information Processing Systems},
 pages = {34752--34764},
 publisher = {Curran Associates, Inc.},
 volume = {36}
}

@inproceedings{LeRoux12,
 author = {Nicolas {Le Roux} and Mark Schmidt and Francis Bach},
 title = {A stochastic gradient method with an exponential convergence rate for finite training sets},
 year = {2012},
 booktitle = {Advances in Neural Information Processing Systems},
  publisher = {Curran Associates, Inc.},
  pages = {2663--2671}
}

@inproceedings{Defazio14,
  author = {Aaron Defazio and Francis Bach and Simon Lacoste-Julien},
  title = {{SAGA}: {{A}} fast incremental gradient method with support for non-strongly convex composite objectives},
  year = {2014},
 booktitle = {Advances in Neural Information Processing Systems},
 volume = {27},
 publisher = {Curran Associates, Inc.},
 pages = {1646--1654}
}

@inproceedings{Johnson13,
  title = {Accelerating stochastic gradient descent using predictive variance reduction},
  booktitle = {Advances in {{Neural Information Processing Systems}}},
  author = {Johnson, Rie and Zhang, Tong},
  year = 2013,
  volume = {26},
  pages = {315--323},
  publisher = {Curran Associates, Inc.},
}

@inproceedings{Vaswani22,
	author = {Sharan Vaswani and Benjamin Dubois-Taine and Reza Babanezhad},
	title = {Towards noise-adaptive, problem-adaptive (accelerated) {SGD}},
	year = {2022},
	booktitle = {Proceedings of the International Conference on Machine Learning},
        series = {{PMLR}},
        pages = {22015--22059},
        volume = {162}
}

@inproceedings{Carmon22,
	author = {Yair Carmon and Oliver Hinder},
	title = {Making {SGD} parameter-free},
	year = {2022},
        series = {{PMLR}},
	booktitle = {Proceedings of the Conference on Learning Theory},
        volume = {178},
        pages = {2360--2389}
}

@inproceedings{Domke19,
  title = {Provable gradient variance guarantees for black-box variational inference},
  booktitle = {Advances in {{Neural Information Processing Systems}}},
  author = {Domke, Justin},
  year = 2019,
  volume = {32},
  pages = {329--338},
  publisher = {Curran Associates, Inc.}
}

@inproceedings{Domke20,
  title = {Provable smoothness guarantees for black-box variational inference},
  booktitle = {Proceedings of the International Conference on Machine Learning},
  author = {Domke, Justin},
  year = 2020,
  series = {{{PMLR}}},
  volume = {119},
  pages = {2587--2596}
}

@inproceedings{Domke23,
  title = {Provable convergence guarantees for black-box variational inference},
  booktitle = {Advances in Neural Information Processing Systems},
  author = {Domke, Justin and Gower, Robert and Garrigos, Guillaume},
  year = 2023,
  volume = {36},
  pages = {66289--66327},
  publisher = {Curran Associates, Inc.}
}

@inproceedings{Kim23,
  title = {On the convergence of black-box variational inference},
  booktitle = {Advances in {{Neural Information Processing Systems}}},
  author = {Kim, Kyurae and Oh, Jisu and Wu, Kaiwen and Ma, Yian and Gardner, Jacob R.},
  year = 2023,
  volume = {36},
  eprint = {2305.15349},
  primaryclass = {cs, eess, math, stat},
  pages = {44615--44657},
  publisher = {Curran Associates Inc.},
}

@article{Duchi11,
  title = {Adaptive subgradient methods for online learning and stochastic optimization},
  author = {Duchi, John and Hazan, Elad and Singer, Yoram},
  year = 2011,
  journal = {Journal of Machine Learning Research},
  volume = {12},
  pages = {2121--2159}
}

@article{Welandawe24,
    author = {Manushi Welandawe and Michael Riis Andersen and Aki Vehtari and Jonathan Huggins},
    title = {A framework for improving the reliability of black-box variational inference},
	journal = {Journal of Machine Learning Research},
	volume = {25},
	number = {219},
	pages = {1--71},
	year = {2024}}

@inproceedings{Kingma15,
  author = {Diederik Kingma and Jimmy Ba},
  title = {Adam: {{A}} method for stochastic optimization},
  booktitle = {Proceedings of the International Conference on Learning Representations},
  year = {2015}}

@inproceedings{Vaswani20,
  author = {Sharan Vaswani and Issam Laradji and Frederik Kunstner and Si Yi Meng and Mark Schmidt and Simon Lacoste-Julien},
  title = {Adaptive gradient methods converge faster with over-parametrization {{(but you should do a line-search)}}},
  booktitle = {Optimization for Machine Learning NeurIPS Workshop},
  year = {2020}}

@inproceedings{Reddi18,
  author = {Sashank Reddi and Satyen Kale and Sanjiv Kumar},
  title = {On the convergence of {Adam} and beyond},
  booktitle = {Proceedings of the International Conference on Learning Representations},
  year = {2018}}

@inproceedings{De17,
  author = {Soham De and Abhay Yadav and David Jacobs and Tom Goldstein},
  title = {{{Big}} {{Batch}} {{SGD}}: {{Automated}} inference using adaptive batch sizes},
  year = {2017},
  booktitle = {Proceedings of the International Conference on Artificial Intelligence and Statistics},
  series = {{PMLR}},
  volume = {52},
  pages = {1504--1513}
}

@inproceedings{Orabona17,
  author = {Francesco Orabona and Tatiana Tommasi},
  title = {Training deep networks without learning rates through coin betting},
  booktitle = {Advances in Neural Information Processing Systems},
  publisher = {Curran Associates, Inc.},
  volume = {30},
  pages = {2160--2170},
  year = {2017}
}

@inproceedings{Defazio23,
  author = {Aaron Defazio and Konstantin Mishchenko},
  title = {Learning-rate-free learning by {{D}}-adaptation},
  year = {2023},
  booktitle = {Proceedings of the International Conference on Machine Learning},
  volume = {202},
  series = {{PMLR}},
  pages = {{7449--7479}}
}

@inproceedings{Ivgi23,
  title = {{{DoG}} is {{SGD}}'s best friend: {{A}} parameter-free dynamic step size schedule},
  shorttitle = {{{DoG}} Is {{SGD}}'s Best Friend},
  booktitle = {Proceedings of the  {{International Conference}} on {{Machine Learning}}},
  author = {Ivgi, Maor and Hinder, Oliver and Carmon, Yair},
  year = 2023,
  series = {{{PMLR}}},
  volume = {202},
  pages = {14465--14499},
}

@inproceedings{Khaled23,
  title = {{{DoWG}} unleashed: {{An}} efficient universal parameter-free gradient descent method},
  shorttitle = {{{DoWG}} Unleashed},
  booktitle = {Advances in {{Neural Information Processing Systems}}},
  author = {Khaled, Ahmed and Mishchenko, Konstantin and Jin, Chi},
  year = 2023,
  volume = {36},
  pages = {6748--6769},
}

@inproceedings{Baydin18,
  author = {{Atilim Gunes} Baydin and Robert Cornish and {David Martinez} Rubio and Mark Schmidt and Frank Wood},
  title = {Online learning rate adaptation with hypergradient descent},
  year = {2018},
  booktitle = {Proceedings of the International Conference on Learning Representations}}

@article{Orabona21,
  author = {Francesco Orabona and D{\'a}vid P{\'a}l},
  title = {Parameter-free stochastic optimization of variationally coherent functions},
  year = {2021},
  journal = {arXiv:2102.00236}}

@inproceedings{Chen23,
  author = {Xiangning Chen and Chen Liang and Da Huang and Esteban Real and Kaiyuan Wang and Yao Liu and Hieu Pham and Xuanyi Dong and Thang Luong and Cho-Jui Hsieh and Yifeng Lu and Quoc Le},
  pages = {49205--49233},
  publisher = {Curran Associates, Inc.},
  volume = {36},
  title = {Symbolic discovery of optimization algorithms},
  year = {2023},
  booktitle = {Advances in Neural Information Processing Systems}

}

@inproceedings{Cutkosky23,
  author = {Ashok Cutkosky and Aaron Defazio and Harsh Mehta},
  title = {Mechanic: {{A}} learning rate tuner},
  year = {2023},
  booktitle = {Advances in Neural Information Processing Systems},
 pages = {47828--47848},
 publisher = {Curran Associates, Inc.},
 volume = {36},
}

@article{Pflug83,
  author = {George Pflug},
  year = {1983},
  title = {On the determination of the step size in stochastic quasigradient methods},
  journal = {IIASA Collaborative Paper CP-83-025}}

@inproceedings{Vaswani19,
  author = {Sharan Vaswani and Aaron Mishkin and Issam Laradji and Mark Schmidt and Gauthier Gidel and Simon Lacoste-Julien},
  title = {Painless stochastic gradient: {{Interpolation}}, line-search, and convergence rates},
  year = {2019},
  volume = {32},
  publisher = {Curran Associates, Inc.},
  booktitle = {Advances in Neural Information Processing Systems},
  pages = {3732--3745}
}

@article{Giordano24,
  author = {Ryan Giordano and Martin Ingram and Tamera Broderick},
  title = {Black box variational inference with a deterministic objective: {{Faster}}, more accurate, and even more black box},
  journal = {Journal of Machine Learning Research},
  volume = {25},
  year = {2024},
  pages = {1--39}}

@incollection{Kim15,
  author = {Sujin Kim and Raghu Pasupathy and Shane Henderson},
  title = {A guide to sample average approximation},
  booktitle = {Handbook of Simulation Optimization},
  publisher = {Springer},
  pages = {207--243},
  year = {2015}}

@inproceedings{Burroni24,
  author = {Javier Burroni and Justin Domke and Daniel Sheldon},
  year = {2024},
  title = {Sample average approximation for black-box variational inference},
  series = {{PMLR}},
  booktitle = {Proceedings of the Conference on Uncertainty in Artificial Intelligence},
  volume = {244},
  pages = {471--498}
}

@misc{viabel,
	author = {{VIABEL Developers}},
	year = {2026},
	title = {{VIABEL}: {{Variational}} inference and approximation bounds that are efficient and lightweight},
	howpublished = {GitHub Repository: \url{https://github.com/jhuggins/viabel}},
	note = {Commit 46dd21a}}

@article{Robbins51,
	author = {Herbert Robbins and Sutton Monro},
	year = {1951},
	title = {A stochastic approximation method},
	journal = {Annals of Mathematical Statistics},
	volume = {22},
	number = {3},
	pages = {400--407}}

@inproceedings{enzyme,
 author = {Moses, William and Churavy, Valentin},
 booktitle = {Advances in Neural Information Processing Systems},
 title = {Instead of Rewriting Foreign Code for Machine Learning, Automatically Synthesize Fast Gradients},
 year = {2020},
 pages = {12472--12485},
 publisher = {Curran Associates, Inc.},
 volume = {33}
}

@inproceedings{Modi23,
  title = {Variational Inference with {{Gaussian}} Score Matching},
  author = {Modi, Chirag and Gower, Robert and Margossian, Charles and Yao, Yuling and Blei, David and Saul, Lawrence},
  year = 2023,
  booktitle = {Advances in Neural Information Processing Systems},
  publisher = {Curran Associates, Inc.},
  volume = {36},
  pages = {29935--29950}
}

@inproceedings{Cai24,
	title = {{{Batch}} and {{Match}}: {{Black}}-box variational inference with a score-based divergence},
	year = {2024},
	author = {Diana Cai and Chirag Modi and Loucas Pillaud-Vivien and Charles Margossian and Robert Gower and David Blei and Lawrence Saul},
  booktitle = {Proceedings International Conference on Machine Learning},
  series = {{PMLR}},
  volume = {235},
  pages = {5258--5297}
  }

@inproceedings{Cai24b,
	title = {{EigenVI}: {{Score}}-based variational inference with orthogonal function expansions},
	year = {2024},
	author = {Diana Cai and Chirag Modi and Charles Margossian and Robert Gower and David Blei and Lawrence Saul},
 pages = {132691--132721},
 publisher = {Curran Associates, Inc.},
  booktitle = {Advances in Neural Information Processing Systems}}

@inproceedings{Gupta18,
  author = {Vineet Gupta and Tomer Koren and Yoram Singer},
  year = {2018},
  title = {Shampoo: {{Preconditioned}} stochastic tensor optimization},
  booktitle = {Proceedings of the International Conference on Machine Learning},
  series = {{PMLR}},
  volume = {80},
  pages = {1842--1850}
}

@misc{Jordan24,
  author       = {Keller Jordan and Yuchen Jin and Vlado Boza and You Jiacheng and
                  Franz Cesista and Laker Newhouse and Jeremy Bernstein},
  title        = {Muon: {{An}} optimizer for hidden layers in neural networks},
  year         = {2024},
  url          = {https://kellerjordan.github.io/posts/muon/}
}

@article{Ba16,
  author = {Jimmy Ba and Jamie Kiros and Geoffrey Hinton},
  year = {2016},
  title = {Layer normalization},
  journal = {arXiv:1607.06450}}

@inproceedings{Kingma14-vae,
  title={Auto-encoding variational {{Bayes}}},
  author={Kingma, Diederik P. and Welling, Max},
  booktitle={Proceedings of the International Conference on Learning Representations},
  year={2014}
}

@article{Margossian22,
  title   = {Flexible and efficient Bayesian pharmacometrics modeling using Stan and Torsten, Part I},
  author  = {Margossian, Charles C. and Zhang, Yi and Gillespie, Bill},
  journal = {CPT: Pharmacometrics \& Systems Pharmacology},
  volume  = {11},
  number  = {9},
  pages   = {1151--1160},
  year    = {2022},
}

@Manual{outbreak,
  title        = {outbreak: Tools for Simulating and Analyzing Epidemic Outbreaks},
  author       = {outbreak package authors},
  year         = {2024},
  note         = {R package},
  url          = {https://CRAN.R-project.org/package=outbreak}
}

@article{Jordan99,
  title = {An introduction to variational methods for graphical models},
  author = {Jordan, Michael I. and Ghahramani, Zoubin and Jaakkola, Tommi S. and Saul, Lawrence K.},
  year = 1999,
  journal = {Machine Learning},
  volume = {37},
  number = {2},
  pages = {183--233},
}

@inproceedings{Titsias14,
  title = {Doubly stochastic variational {{Bayes}} for non-conjugate inference},
  booktitle = {Proceedings of the {{International Conference}} on {{Machine Learning}}},
  author = {Titsias, Michalis and {L{\'a}zaro-Gredilla}, Miguel},
  year = 2014,
  series = {{{PMLR}}},
  volume = {32},
  pages = {1971--1979},
}

@inproceedings{Ranganath14,
  title = {Black box variational inference},
  booktitle = {Proceedings of the International Conference on Artificial Intelligence and Statistics},
  author = {Ranganath, Rajesh and Gerrish, Sean and Blei, David},
  year = 2014,
  series = {{{PMLR}}},
  volume = {33},
  pages = {814--822},
}

@article{Wingate13,
  title = {Automated variational inference in probabilistic programming},
  author = {Wingate, David and Weber, Theophane},
  year = 2013,
  journal = {arXiv:1301.1299},
}

@article{Kucukelbir17,
  title = {Automatic differentiation variational inference},
  author = {Kucukelbir, Alp and Tran, Dustin and Ranganath, Rajesh and Gelman, Andrew and Blei, David M.},
  year = 2017,
  journal = {Journal of Machine Learning Research},
  volume = {18},
  number = {14},
  pages = {1--45}
}

@article{Blei17,
  title = {Variational inference: {{A}} review for statisticians},
  shorttitle = {Variational {{Inference}}},
  author = {Blei, David M. and Kucukelbir, Alp and McAuliffe, Jon D.},
  year = 2017,
  journal = {Journal of the American Statistical Association},
  volume = {112},
  number = {518},
  pages = {859--877},
}

@inproceedings{Hinton93,
  title = {Keeping the neural networks simple by minimizing the description length of the weights},
  booktitle = {Proceedings of the Annual Conference on {{Computational}} Learning Theory},
  author = {Hinton, Geoffrey E. and {van Camp}, Drew},
  year = 1993,
  pages = {5--13},
  publisher = {ACM Press},
}

@article{Peterson89,
  title = {Explorations of the mean field theory learning algorithm},
  author = {Peterson, Carsten and Hartman, Eric},
  year = 1989,
  journal = {Neural Networks},
  volume = {2},
  number = {6},
  pages = {475--494},
}

@inproceedings{Minka01,
  title = {Expectation propagation for approximate {{Bayesian}} inference},
  booktitle = {Proceedings of the Conference on Uncertainty in Artificial Intelligence},
  author = {Minka, Thomas P.},
  year = 2001,
  pages = {362--369},
  publisher = {Morgan Kaufmann Publishers Inc.},
  address = {San Francisco, CA, USA},
}

@incollection{Bottou99,
  title = {On-line learning and stochastic approximations},
  booktitle = {On-{{Line Learning}} in {{Neural Networks}}},
  author = {Bottou, L{\'e}on},
  year = 1999,
  pages = {9--42},
  editor = {David Saad},
  publisher = {Cambridge University Press},
}

@article{Bottou18,
  title = {Optimization methods for large-scale machine learning},
  author = {Bottou, L{\'e}on and Curtis, Frank E. and Nocedal, Jorge},
  year = 2018,
  journal = {SIAM Review},
  volume = {60},
  number = {2},
  pages = {223--311},
}

@inproceedings{Rezende14,
  title = {Stochastic Backpropagation and Approximate Inference in Deep Generative Models},
  booktitle = {Proceedings of the {{International Conference}} on {{Machine Learning}}},
  author = {Rezende, Danilo Jimenez and Mohamed, Shakir and Wierstra, Daan},
  year = 2014,
  series = {{{PMLR}}},
  volume = {32},
  pages = {1278--1286},
}

@article{Mohamed20,
  title = {Monte {{Carlo}} Gradient Estimation in Machine Learning},
  author = {Mohamed, Shakir and Rosca, Mihaela and Figurnov, Michael and Mnih, Andriy},
  year = 2020,
  journal = {Journal of Machine Learning Research},
  volume = {21},
  number = {132},
  pages = {1--62},
}

@inproceedings{Graves11,
  title = {Practical variational inference for neural networks},
  booktitle = {Advances in {{Neural Information Processing Systems}}},
  author = {Graves, Alex},
  year = 2011,
  volume = {24},
  pages = {2348--2356},
  publisher = {Curran Associates, Inc.},
}

@article{Saul96,
  title = {Mean field theory for sigmoid belief networks},
  author = {Saul, Lawrence K. and Jaakkola, Tommi and Jordan, Michael I.},
  year = 1996,
  journal = {Journal of Artificial Intelligence Research},
  volume = {4},
  pages = {61--76},
}

@inproceedings{Schmidt21,
  title = {Descending through a crowded valley---{Benchmarking} deep learning optimizers},
  booktitle = {Proceedings of the {{International Conference}} on {{Machine Learning}}},
  author = {Schmidt, Robin M. and Schneider, Frank and Hennig, Philipp},
  year = 2021,
  series = {{{PMLR}}},
  volume = {139},
  pages = {9367--9376},
}

@inproceedings{Yao18,
  title = {Yes, but did it work?: {{Evaluating}} variational inference},
  shorttitle = {Yes, but Did It Work?},
  booktitle = {Proceedings of the {{International Conference}} on {{Machine Learning}}},
  author = {Yao, Yuling and Vehtari, Aki and Simpson, Daniel and Gelman, Andrew},
  year = 2018,
  series = {{{PMLR}}},
  volume = {80},
  pages = {5581--5590},
}

@inproceedings{Dhaka20,
  title = {Robust, accurate stochastic optimization for variational inference},
  booktitle = {Advances in {{Neural Information Processing Systems}}},
  author = {Dhaka, Akash Kumar and Catalina, Alejandro and Andersen, Michael R and Magnusson, M{\aa}ns and Huggins, Jonathan and Vehtari, Aki},
  year = 2020,
  volume = {33},
  pages = {10961--10973},
  publisher = {Curran Associates, Inc.},
}

@inproceedings{Huggins20,
  title = {Validated variational inference via practical posterior error bounds},
  booktitle = {Proceedings of the {{Twenty Third International Conference}} on {{Artificial Intelligence}} and {{Statistics}}},
  author = {Huggins, Jonathan and Kasprzak, Mikolaj and Campbell, Trevor and Broderick, Tamara},
  year = 2020,
  series = {{{PMLR}}},
  volume = {108},
  pages = {1792--1802},
}

@inproceedings{Dhaka21,
  title = {Challenges and opportunities in high-dimensional variational inference},
  booktitle = {Advances in {{Neural Information Processing Systems}}},
  author = {Dhaka, Akash Kumar and Catalina, Alejandro and Welandawe, Manushi and Andersen, Michael R. and Huggins, Jonathan and Vehtari, Aki},
  year = 2021,
  volume = {34},
  pages = {7787--7798},
  publisher = {Curran Associates, Inc.}
}

@inproceedings{Geffner21a,
  title = {On the difficulty of unbiased alpha divergence minimization},
  booktitle = {Proceedings of the International Conference on Machine Learning},
  author = {Geffner, Tomas and Domke, Justin},
  year = 2021,
  series = {{{PMLR}}},
  volume = {139},
  pages = {3650--3659},
}

@inproceedings{Geffner21b,
  title = {Empirical evaluation of biased methods for alpha divergence minimization},
  booktitle = {Proceedings of the Symposium on Advances in Approximate Bayesian Inference},
  author = {Geffner, Tomas and Domke, Justin},
  year = {2021}
}

@article{Dayan95,
  title = {The {{Helmholtz}} Machine},
  author = {Dayan, Peter and Hinton, Geoffrey E. and Neal, Radford M. and Zemel, Richard S.},
  year = 1995,
  journal = {Neural Computation},
  volume = {7},
  number = {5},
  pages = {889--904},
}

@book{Robert04,
  title = {Monte {{Carlo Statistical Methods}}},
  author = {Robert, Christian P. and Casella, George},
  year = 2004,
  series = {Springer {{Texts}} in {{Statistics}}},
  publisher = {Springer},
}

@article{Roualdes23,
  title = {{{BridgeStan}}: {{Efficient}} in-memory access to the methods of a {{Stan}} model},
  shorttitle = {{{BridgeStan}}},
  author = {Roualdes, Edward A. and Ward, Brian and Carpenter, Bob and Seyboldt, Adrian and Axen, Seth D.},
  year = 2023,
  journal = {Journal of Open Source Software},
  volume = {8},
  number = {87},
  pages = {5236},
}

@article{Bezanson17,
  title = {Julia: {{A}} fresh approach to numerical computing},
  author = {Bezanson, Jeff and Edelman, Alan and Karpinski, Stefan and Shah, Viral B},
  year = 2017,
  journal = {SIAM review},
  volume = {59},
  number = {1},
  pages = {65--98}
}

@article{Khan23,
  title = {The {{Bayesian}} learning rule},
  author = {Khan, Mohammad Emtiyaz and Rue, H{\aa}vard},
  year = 2023,
  journal = {Journal of Machine Learning Research},
  volume = {24},
  number = {281},
  pages = {1--46},
}

@article{Kumar25,
  title = {Optimization guarantees for square-root natural-gradient variational inference},
  author = {Kumar, Navish and M{\"o}llenhoff, Thomas and Khan, Mohammad Emtiyaz and Lucchi, Aurelien},
  year = 2025,
  journal = {Transactions on Machine Learning Research},
}

@inproceedings{Lin19,
  title = {Fast and simple natural-gradient variational inference with mixture of exponential-family approximations},
  booktitle = {Proceedings of the {{International Conference}} on {{Machine Learning}}},
  author = {Lin, Wu and Khan, Mohammad Emtiyaz and Schmidt, Mark},
  year = 2019,
  series = {{{PMLR}}},
  volume = {97},
  pages = {3992--4002},
}

@inproceedings{Lambert22,
  title = {Variational inference via {{Wasserstein}} gradient flows},
  booktitle = {Advances in {{Neural Information Processing Systems}}},
  author = {Lambert, Marc and Chewi, Sinho and Bach, Francis and Bonnabel, Silv{\`e}re and Rigollet, Philippe},
  year = 2022,
  volume = {35},
  pages = {14434--14447},
  publisher = {Curran Associates, Inc.},
}

@inproceedings{Diao23,
  title = {Forward-backward {{Gaussian}} variational inference via {{JKO}} in the {{Bures-Wasserstein}} space},
  booktitle = {Proceedings of the {{International Conference}} on {{Machine Learning}}},
  author = {Diao, Michael Ziyang and Balasubramanian, Krishna and Chewi, Sinho and Salim, Adil},
  year = 2023,
  series = {{{PMLR}}},
  volume = {202},
  pages = {7960--7991},
}

@article{Huszar17,
title={Variational inference using implicit distributions}, 
author={Ferenc Huszár},
year={2017},
journal={arXiv:1702.08235},
}

@inproceedings{Yu23,
title={Semi-implicit variational inference via score matching},
author={Longlin Yu and Cheng Zhang},
booktitle={Proceedings of the International Conference on Learning Representations },
year={2023},
}

@inproceedings{Ma18,
  title = {The power of interpolation: {{Understanding}} the effectiveness of {{SGD}} in modern over-parametrized learning},
  shorttitle = {The Power of Interpolation},
  booktitle = {Proceedings of the {{International Conference}} on {{Machine Learning}}},
  author = {Ma, Siyuan and Bassily, Raef and Belkin, Mikhail},
  year = 2018,
  series = {{{PMLR}}},
  volume = {80},
  pages = {3325--3334},
}

@inproceedings{Kim24,
  title = {Linear convergence of black-box variational inference: {{Should}} we stick the landing?},
  booktitle = {Proceedings of the {{International Conference}} on {{Artificial Intelligence}} and {{Statistics}}},
  author = {Kim, Kyurae and Ma, Yian and Gardner, Jacob R.},
  year = 2024,
  series = {{{PMLR}}},
  volume = {238},
  pages = {235--243},
}

@article{Ho83,
  title = {Perturbation analysis and optimization of queueing networks},
  author = {Ho, Y. C. and Cao, X.},
  year = 1983,
  journal = {Journal of Optimization Theory and Applications},
  volume = {40},
  number = {4},
  pages = {559--582},
}

@article{Rubinstein92,
  title = {Sensitivity analysis of discrete event systems by the ``push out'' method},
  author = {Rubinstein, Reuven Y.},
  year = 1992,
  journal = {Annals of Operations Research},
  volume = {39},
  number = {1},
  pages = {229--250},
}

@inproceedings{Burda15,
  title = {Importance weighted autoencoders},
  booktitle = {Proceedings of the {{International}} Conference on Learning Representations},
  author = {Burda, Yuri and Grosse, Roger and Salakhutdinov, Ruslan},
  year = 2015
}

@inproceedings{Domke18,
  title = {Importance weighting and variational inference},
  booktitle = {Advances in {{Neural Information Processing Systems}}},
  author = {Domke, Justin and Sheldon, Daniel R},
  year = 2018,
  volume = {31},
  publisher = {Curran Associates, Inc.},
  pages = {4470--4479}
}

@inproceedings{Roeder17,
  title = {Sticking the landing: {{Simple}}, lower-variance gradient estimators for variational inference},
  booktitle = {Advances in {{Neural Information Processing Systems}}},
  author = {Roeder, Geoffrey and Wu, Yuhuai and Duvenaud, David K},
  year = 2017,
  volume = {30},
  pages = {6928--6937},
  publisher = {Curran Associates, Inc.}
}

@inproceedings{Rezende15,
  title = {Variational Inference with Normalizing Flows},
  booktitle = {Proceedings of the {{International Conference}} on {{Machine Learning}}},
  author = {Rezende, Danilo and Mohamed, Shakir},
  year = 2015,
  series = {{{PMLR}}},
  volume = {37},
  pages = {1530--1538},
}

@inproceedings{Hernandezlobato16,
  title = {Black-Box alpha divergence minimization},
  booktitle = {Proceedings of the International Conference on Machine Learning},
  author = {{Hernandez-Lobato}, Jose and Li, Yingzhen and Rowland, Mark and Bui, Thang and {Hernandez-Lobato}, Daniel and Turner, Richard},
  year = 2016,
  series = {{{PMLR}}},
  volume = {48},
  pages = {1511--1520},
}

@inproceedings{Dieng17,
  title = {Variational inference via {$\chi$}-upper bound minimization},
  booktitle = {Advances in {{Neural Information Processing Systems}}},
  author = {Dieng, Adji Bousso and Tran, Dustin and Ranganath, Rajesh and Paisley, John and Blei, David},
  year = 2017,
  volume = {30},
  pages = {2729--2738},
  publisher = {Curran Associates, Inc.},
}

@inproceedings{Geffner21,
  title = {{{MCMC}} variational inference via uncorrected {{Hamiltonian}} annealing},
  booktitle = {Advances in Neural Information Processing Systems},
  author = {Geffner, Tomas and Domke, Justin},
  year = 2021,
  volume = {34},
  pages = {639--651},
  publisher = {Curran Associates, Inc.},
}

@inproceedings{Zhang21,
  title = {Differentiable annealed importance sampling and the perils of gradient noise},
  booktitle = {Advances in Neural Information Processing Systems},
  author = {Zhang, Guodong and Hsu, Kyle and Li, Jianing and Finn, Chelsea and Grosse, Roger B},
  year = 2021,
  volume = {34},
  pages = {19398--19410},
  publisher = {Curran Associates, Inc.},
}

@inproceedings{Domke19-dividecouple,
 author = {Domke, Justin and Sheldon, Daniel R},
 booktitle = {Advances in Neural Information Processing Systems},
 publisher = {Curran Associates, Inc.},
 title = {{{Divide}} and {{Couple}}: using {{Monte}} {{Carlo}} variational objectives for posterior approximation},
 volume = {32},
 year = {2019},
 pages = {339--349},
}

@inproceedings{Buchholz18,
  title = {Quasi-{{Monte Carlo}} variational inference},
  booktitle = {Proceedings of the {{International Conference}} on {{Machine Learning}}},
  author = {Buchholz, Alexander and Wenzel, Florian and Mandt, Stephan},
  year = 2018,
  series = {{{PMLR}}},
  volume = {80},
  pages = {668--677},
}

@article{Fujisawa21,
  title = {Multilevel {{Monte Carlo}} variational inference},
  author = {Fujisawa, Masahiro and Sato, Issei},
  year = 2021,
  journal = {Journal of Machine Learning Research},
  volume = {22},
  number = {278},
  pages = {1--44},
}

@article{Ong18,
  title = {Gaussian Variational Approximation with a Factor Covariance Structure},
  author = {Ong, Victor M.-H. and Nott, David J. and Smith, Michael S.},
  year = 2018,
  journal = {Journal of Computational and Graphical Statistics},
  volume = {27},
  number = {3},
  pages = {465--478},
  publisher = {Taylor \& Francis},
}

@inproceedings{Ranganath16,
  title = {Hierarchical variational models},
  booktitle = {Proceedings of the {{International Conference}} on {{Machine Learning}}},
  author = {Ranganath, Rajesh and Tran, Dustin and Blei, David},
  year = 2016,
  series = {{{PMLR}}},
  volume = {48},
  pages = {324--333},
}

@InProceedings{Titsias19,
  title = 	 {Unbiased implicit variational inference},
  author =       {Titsias, Michalis K. and Ruiz, Francisco},
  booktitle = 	 {Proceedings of the International Conference on Artificial Intelligence and Statistics},
  pages = 	 {167--176},
  year = 	 {2019},
  volume = 	 {89},
  series =    {{PMLR}},
}

@inproceedings{Salimans15,
  title = {Markov Chain {{Monte Carlo}} and Variational Inference: {{Bridging}} the Gap},
  booktitle = {Proceedings of the {{International Conference}} on {{Machine Learning}}},
  author = {Salimans, Tim and Kingma, Diederik and Welling, Max},
  year = 2015,
  series = {{{PMLR}}},
  volume = {37},
  pages = {1218--1226},
}

@article{Tan18,
  title = {Gaussian variational approximation with sparse precision matrices},
  author = {Tan, Linda S. L. and Nott, David J.},
  year = 2018,
  journal = {Statistics and Computing},
  volume = {28},
  number = {2},
  pages = {259--275},
}

@inproceedings{Li16,
  title = {R\'enyi divergence variational inference},
  booktitle = {Advances in {{Neural Information Processing Systems}}},
  author = {Li, Yingzhen and Turner, Richard E},
  year = 2016,
  volume = {29},
  publisher = {Curran Associates, Inc.},
  pages = {1073--1081}
}

@article{Tan25,
  title = {Analytic natural gradient updates for {{Cholesky}} factor in {{Gaussian}} variational approximation},
  author = {Tan, Linda S L},
  year = 2025,
  journal = {Journal of the Royal Statistical Society Series B: Statistical Methodology},
  volume = {87},
  number = {4},
  pages = {930--956},
}

@article{Gelman92,
  author = {Andrew Gelman and Donald Rubin},
  year = {1992},
  title = {Inference from iterative simulation using multiple sequences},
  journal = {Statistical Science},
  volume = {7},
  number = {4},
  pages = {457--511}}

@inproceedings{He16,
author = {He, Kaiming and Zhang, Xiangyu and Ren, Shaoqing and Sun, Jian},
title = {Deep residual learning for image recognition},
booktitle = {Proceedings of the IEEE Conference on Computer Vision and Pattern Recognition},
month = {June},
year = {2016},
pages={770-778}
}

@inproceedings{Nair10,
author = {Nair, Vinod and Hinton, Geoffrey E.},
title = {Rectified linear units improve restricted {{Boltzmann}} machines},
year = {2010},
publisher = {Omnipress},
address = {Madison, WI, USA},
booktitle = {Proceedings of the International Conference on International Conference on Machine Learning},
pages = {807–814},
numpages = {8},
series = {ICML}
}

\clearpage

\appendix

\section{Subsampled PosteriorDB}\label{sec:data-subsampled-pdb}

As part of our benchmark problem suite, we have re-implemented 127 out of the 146 posterior
distributions from \texttt{posteriordb} \cite{posteriordb} to enable subsampled estimates
of sums in the target log probability density, which typically (but not always) arise due to 
the presence of conditionally independent data terms. The main challenge in doing so is that Stan accepts 
only real-valued vectors as arguments to the target (gradient) log probability density function.
\cref{lst:subsampledstan} demonstrates a simple example of the general technique we used to incorporate
subsampling in the Stan code on the \texttt{earn\_height} model. In particular, 
we include a new real-valued parameter \texttt{SUBIDX} in the model, and within the model code we loop
over all data indices and add to the target only that corresponding to \texttt{SUBIDX} (because \texttt{SUBIDX}
is real-valued, we check that $i-0.5 \leq \texttt{SUBIDX} \leq i+0.5$ to avoid floating point issues).
When evaluating the target log probability density at a parameter value $x\in\reals^d$, we randomly draw $i \dist \Unif\{1,\dots, N\}$
(where $N$ is the number of terms in the subsampled model),
pass $(x, i) \in \reals^{d+1}$ to the Stan model, and receive unbiased log density and gradient 
evaluation $N \log p(x, i)\in\reals $, $(N \grad \log p(x, i), 0)\in\reals^{d+1}$. We discard
the last index in the $(d+1)$-dimensional gradient output, which is always 0.

Note that a key limitation of this technique is that it requires running a \texttt{for} loop to search for the term that matches \texttt{SUBIDX}.
Therefore we do not expect to obtain the full benefit of data-subsampling one would expect in practice.
However, for the purposes of this paper, this is unimportant; all algorithms applied to subsampled problems
experience the same cost-per-evaluation (regardless of whether it is much faster than the full-data evaluation).
We do not recommend using this strategy if one wants to compare the performance of optimization when using
subsampled log probability density evaluations versus when using full log probability density evaluations.
Note that there are other potential strategies for adding indexing to the Stan model that avoid the \texttt{for} loop  
(e.g., adding the index to the \texttt{data\{...\}} block), but without modifying 
Stan and/or BridgeStan, we found these to be substantially
slower in practice due to the inability to efficiently modify the index variable at runtime.

\begin{figure}[t]
\caption{Stan code for the original \texttt{earn\_height} model \cite{posteriordb},
followed by Stan code for the subsampled \texttt{earn\_height\_subsampled} model. Note the additional
parameter \texttt{SUBIDX} that is appended to the state, treated as an extra dimension 
in the BridgeStan input argument, and then used to select just one of the data log-likelihood terms to
include in the \texttt{target} log probability density.}\label{lst:subsampledstan}
\begin{lstlisting}
// Original model code
data {
  int<lower=0> N;
  vector[N] earn;
  vector[N] height;
}
parameters {
  vector[2] beta;
  real<lower=0> sigma;
}
model {
  earn ~ normal(beta[1]+beta[2]*height, sigma);
}

// New subsampled model code
data {
  int<lower=0> N;
  vector[N] earn;
  vector[N] height;
}
parameters {
  vector[2] beta;
  real<lower=0> sigma;
  real SUBIDX;
}
model {
for (i in 1:N){
if (i-0.5 <= SUBIDX && i+0.5 >= SUBIDX){
  target += N*normal_lpdf(earn[i] | beta[1]+beta[2]*height[i], sigma);
  break;
}
}
}
\end{lstlisting}
\end{figure}

\clearpage

\section{Neural Network Architecture}\label{sec:neuralnets}

For the variational autoencoder (VAE) problems, the neural network 
structure we use for both the encoder and the decoder is identical.
Here we describe and provide pseudocode for the decoder, \textit{i.e.}, the network
that receives $Z \in \reals^d$ and outputs a mean $\mu_\lambda(Z)$ 
and covariance matrix $\Sigma_\lambda(Z)$ to be used in the augmented variational
family from \cref{eq:augvaefam}.
Pseudocode for the neural network is provided in \cref{alg:nn}.
First, $Z \in \reals^d$ is padded with zeros: $y \gets (Z, 0)\in\reals^{2d}$.
Then we compose layers with a low-rank ($2d\times d_{\text{int}}$, where $d_{\text{int}}$ is a small fixed value)
affine transformation to avoid extreme parameter dimensions $A_n B_n y + b_n$, 
a rectified linear unit (RELU) nonlinear activation, a residual skip connection, and finally a
a form of layer normalization \cite{Ba16}. 
The final output layer involves a full-rank affine transformation.

The parameter dimension of the network is computed as follows. 
Suppose the original target log density has a parameter of dimension $d\in\nats$. 
Since each $A_n,B_n \in \reals^{(2d)\times d_{\text{int}}}$ and $b_n \in \reals^{2d}$, 
each internal layer in the neural network has $2d + 4d d_{\text{int}}$ parameters, and there
are $n_{\text{layers}}$ of these layers total. At the output layer we have $A \in\reals^{(2d)\times(2d)}$
and $b \in \reals^{2d}$, resulting in another $4d^2+2d$ parameters.
Therefore each neural network has $n_{\text{layers}}(2d + 4d d_{\text{int}}) + 4d^2 + 2d$ parameters.
Since there are $2$ such neural networks---one for the encoder, and one for the decoder---and $n_{\text{layers}}=d_{\text{int}}=10$,
we have a total of
\[
2( 4d^2 + 2d + 10(40d + 2d) ) = 8d^2 + 4d + 800d + 40d = 8d^2 + 844d
\]
tunable parameters in the VAE inference problems.

\balg
\caption{Neural network structure for the decoder $(\mu_\lambda, \Sigma_\lambda)(Z)$.
The output is a $2d$-dimensional vector, where the first $d$ components are used as $\mu_{\lambda}(Z)$
and the latter $d$ components are used as 
$\sigma_{\lambda,1}(Z), \dots, \sigma_{\lambda,d}(Z)$ (which may have positive or negative sign),
where
$\Sigma_\lambda(Z) = \diag \sigma^2_{\lambda,1}(Z), \dots, \sigma^2_{\lambda,d}(Z)$.
We use precisely the same structure for the encoder $(m_\lambda, S_\lambda)(X)$.
In our experiments we use $d_{\text{int}} = n_{\text{layers}} = 10$.}\label{alg:nn}
\balgc
\Require Input noise $\epsilon \in \reals^{d}$, 
internal dimension $d_{\text{int}}$, number of layers $n_{\text{layers}}$,
parameters $\lambda = (A, b, A_1, B_1, b_1, \dots, A_{n_\text{layers}}, B_{n_\text{layers}}, b_{n_\text{layers}})$
\LineComment{Pad $\epsilon$ with zeros to get the neural network state vector $y$}
\State $y \gets (\epsilon, \texttt{zeros}(d)) \in \reals^{2d}$
\For{$n=1,\dots, n_\text{layers}$}
\LineComment[1]{ResNet layer with RELU activation}
\LineComment[1]{$A_n\in \reals^{2d\times d_{\text{int}}}$, $B_n\in\reals^{d_\text{int}\times 2d}$, $b_n\in\reals^{2d}$}
\State $y \gets \max\lt(0, A_n B_n y + b_n\rt) + y$ 
\LineComment{Normalization layer ($1$ is the $\reals^{2d}$ vector of all $1$ entries)}
\State $\mu \gets 1^Ty/(2d)$
\State $y \gets \frac{y-\mu 1}{\|y-\mu 1\|}$
\EndFor
\LineComment{Final linear layer with $A\in\reals^{2d\times 2d}, b\in\reals^{2d}$}
\State $y \gets Ay + b$
\State $\mu \gets y_{1:d}$, $\Sigma \gets \lt(\diag y_{d+1:2d}\rt)^2$
\State \Return $\mu, \Sigma$
\ealgc
\ealg

\clearpage

\section{Additional Algorithms}\label{sec:additionalalgs}

\subsection{Sample Average Approximation Methods}\label{sec:saa}

The full detailed pseudocode for our proposed sample average approximation (SAA) method is provided
in \cref{alg:saa}. This is a generic template for all of the specific SAA instantiations that appear
in this work (SAA, SAANA, SAACG, SAALBFGS). First we obtain a gradient estimate, and use it to compute
a search direction with the current batch size $n$. Next, we double the batch size $n$ until that search 
direction agrees reasonably well with the search direction at batch size $2n$.
Then we perform an Armijo line search, take the step, and increase (``reset'') the step size.
To implement variants of this method, one needs to specify the \texttt{SearchDirection}, \texttt{InitializeMemory},
and \texttt{UpdateMemory} functions. For example, for basic gradient descent (\cref{alg:saabasic}), the \texttt{SearchDirection} function
simply returns the gradient, and the \texttt{InitializeMemory} and \texttt{UpdateMemory} functions do nothing.
For conjugate gradient (\cref{alg:saacg}), we store and incrementally update the conjugate gradient search direction.

\balg[t]
\caption{One step of the Sample Average Approximation (SAA), with a generic \texttt{SearchDirection} functions. 
By specifying different \texttt{SearchDirection} and \texttt{UpdateMemory} functions, one can implement SAA, SAANA, SAACG, SAALBFGS, and other variants.
Variants for SAA and SAACG are shown in \cref{alg:saabasic,alg:saacg}.
Note that this pseudocode presents a 
na\"ive and inefficient implementation
for the purposes of clarity; an efficient implementation involves a significant amount of caching to avoid recomputing expensive
function and gradient evaluations. Also note that the same seed value $s$ appears many times throughout the code; it is essential to keep
re-using the same seed value to guarantee a single, growing batch of draws.}\label{alg:saa}
\balgc
\Require current state $x$, current step size $\gamma$, current batch size $n$, memory/stored statistics $m$, PRNG seed $s$,
Armijo constant $\eta = 0.5$, line search factor $c=2$
\LineComment{Draw the gradient estimate at $x$ with batch size $n$ and PRNG initialized at the same seed each step}
\State $g \gets \texttt{GradientEstimate}(x, n, s)$
\State $g_2 \gets \texttt{GradientEstimate}(x, 2n, s)$
\LineComment{Compute the search direction for batch sizes $n$, $2n$} 
\State $p \gets \texttt{SearchDirection}(x, g_1, m)$
\State $p_2 \gets \texttt{SearchDirection}(x, g_2, m)$
\LineComment{Increase $n$ until the search direction at batch size $n$ agrees with that of $2n$}
\While{$\|p\| \leq (1/2)\|p_2\|$ or $p^Tp_2 < 0$}
\LineComment{Double the batch size}
\State $n \gets 2n$
\LineComment{Re-draw the gradient estimate at $x$ with batch size $n$, PRNG seed $s$}
\State $g \gets \texttt{GradientEstimate}(x, n, s)$
\State $g_2 \gets \texttt{GradientEstimate}(x, 2n, s)$
\LineComment{Re-compute the search direction for batch sizes $n$, $2n$} 
\State $p \gets \texttt{SearchDirection}(x, g, m)$
\State $p_2 \gets \texttt{SearchDirection}(x, g_2, m)$
\EndWhile
\State $f \gets \texttt{FunctionEstimate}(x, n, s)$
\State $f' \gets \texttt{FunctionEstimate}(x-\gamma p, n, s)$
\LineComment{Armijo line search}
\While{$f' > f - \eta\gamma p^Tg$}
\State $\gamma \gets \gamma/c$
\State $f' \gets \texttt{FunctionEstimate}(x-\gamma p, n, s)$
\EndWhile
\LineComment{Update the algorithm memory}
\State $m\gets \texttt{UpdateMemory}(x, \gamma, p, g, m)$
\LineComment{Take the step}
\State $x'\gets x-\gamma p$
\LineComment{Reset the step size}
\State $\gamma \gets \gamma c$
\State \Return $x', \gamma, n, m$
\ealgc
\ealg

\balg[t]
\caption{Search direction, memory initialize, and memory update computations for basic gradient descent (SAA). 
Gradient descent uses $g$ as the search
direction and does not require the storage and update of any auxiliary statistics.}\label{alg:saabasic}
\raggedright
$\texttt{SearchDirection}(x,g,m)$
\balgc
\Require state $x$, gradient estimate $g$, memory $m$
\State \Return $g$
\ealgc
\hphantom{x}\newline
$\texttt{InitializeMemory}(d)$
\balgc
\Require dimension of the state variable $d$
\State \Return null
\ealgc
\hphantom{x}\newline
$\texttt{UpdateMemory}(x,\gamma,p,g,m)$
\balgc
\Require state $x$, direction $p$, gradient estimate $g$, memory $m$
\State \Return null
\ealgc
\ealg

\balg[t]
\caption{Search direction, memory initialize, and memory update computations for nonlinear conjugate gradient (SAACG).}\label{alg:saacg}
\raggedright
$\texttt{SearchDirection}(x,g,m)$
\balgc
\Require state $x$, gradient estimate $g$, memory $m = (p_\text{prev}, g_\text{prev})$
\State $\beta \gets \frac{\max(0, g^T(g - g_{\text{prev}}))}{\|g_{\text{prev}}\|^2}$ \Comment{Polak-Ribi\`ere update}
\State \Return $g + \beta p_{\text{prev}}$
\ealgc
\hphantom{x}\newline
$\texttt{InitializeMemory}(d)$
\balgc
\Require dimension of the state variable $d$
\State $p_{\text{prev}} \gets \texttt{zeros}(d)$
\State $g_{\text{prev}} \gets \texttt{ones}(d)$ \Comment{Anything nonzero for $g_{\text{prev}}$ will work}
\State \Return $p_{\text{prev}}$, $g_{\text{prev}}$
\ealgc
\hphantom{x}\newline
$\texttt{UpdateMemory}(x,\gamma,p,g,m)$
\balgc
\Require state $x$, direction $p$, gradient estimate $g$, memory $m = (p_\text{prev}, g_\text{prev})$
\State $p_{\text{prev}} \gets p$
\State $g_{\text{prev}} \gets g$
\State \Return $(p_{\text{prev}}, g_{\text{prev}})$
\ealgc
\ealg

\clearpage

% !TEX root =  main.tex

\subsection{Streaming RABVI}\label{sec:srabvi}

The RABVI meta-algorithm \cite{Welandawe24} adaptively reduces the step size
$\gamma$ of a base stochastic optimizer by monitoring convergence diagnostics
applied to the optimization iterates (in the present context, the optimization iterates are the 
variational parameters).  In detail, RABVI applies three convergence diagnostics,
originally developed for iterates generated by Markov chain Monte Carlo (MCMC):
the potential scale reduction factor $\shR$ \cite{Gelman92,Vehtari21}, the
effective sample size (ESS), and the Monte Carlo standard error (MCSE).
In line with standard practices of MCMC, RABVI first accumulates iterates until the $\shR$ 
drops below $1.1$---with $\shR \approx 1$ indicating convergence to a stationary distribution under non-adversarial conditions---after examining several possible window sizes.
RABVI then continues accumulating iterates until a desired MCSE and ESS are achieved across all coordinates.
At this point, it declares convergence, reports the current Monte Carlo estimate, and reduces step size.
Crucially, however, RABVI stores every iterate produced within a step-size phase,
which becomes prohibitive when $d$ is large. 
Furthermore, the original algorithm is designed to be run to convergence, 
not as an anytime algorithm. 

To address these issues, we develop \emph{streaming RABVI} (SRABVI; \cref{alg:srabvi}), a low-memory variant
that replaces storage of the full iterate trace with tracking a hierarchy of batch
sufficient statistics.
SRABVI computes all three diagnostics directly from those summaries.
During batch construction, iterates are accumulated using Welford's online
algorithm \cite{Welford62}, retaining only a coordinate-wise mean, variance,
and count. Completed batches are stored in a binary-counter hierarchy that
keeps at most two batches at each ``merge level'': when a third batch
arrives at a level, the two oldest are combined (using the numerically
stable parallel-variance formula of \citet{Chan83}) and promoted to the
next level, with carry propagation upward. 
Given an initial minimum batch size of $\Wmin$, after $k$ iterates there are
at most $O(\log(k/\Wmin))$ stored batches.
The compressed batch summaries are then used to
compute split-chain $\shR$ \cite{Vehtari21} and a batch-means ESS and MCSE
estimator \cite{Flegal10,Vats19,Liu22}, both generalized to handle the
unequal batch sizes produced by the merge hierarchy. We also replace the
fixed-ratio step-size rule $\gamma \gets \rho\gamma$ of the original RABVI
with a quadratic rule $\gamma \gets \gamma^2$, eliminating the hyperparameter
$\rho$ and yielding a doubly exponential schedule that is justified by
balancing the gradient noise at the new step size against the optimization
bias at the previous one.

A full Julia implementation of streaming RABVI is available online at \url{https://github.com/trevorcampbell/defaultvi}.

\balg[t]
\caption{One step of streaming RABVI (SRABVI). The base optimizer maintains
state $x$ and uses fixed step size $\gamma$. SRABVI updates $x$ and $\gamma$
and reports a running iterate average $\bar x$. The sub-procedures
\texttt{Rebalance}, \texttt{WindowedRhat}, and \texttt{StreamingMCSE} are
described in the text and operate entirely on the compressed batch
summaries.}\label{alg:srabvi}
\balgc
\Require base optimizer with state $x$ and step size $\gamma$,
initial MCSE tolerance $\epsilon$, initial minimum batch size $\Wmin$,
ESS threshold $\ESS_{\min}$
\Require persistent state: batch list $\mathcal{L}$, Welford accumulator
$(\mu, M_2, c)$ and target batch size $b$, within-phase iterate count $n$,
iterate-average sum $S$ and count $k$, $\shR$-convergence marker $k_\mathrm{conv}$
and recheck window $W_\mathrm{chk}$, previous-phase summary
$(\bar x_\mathrm{prev}, n_\mathrm{prev}, \gamma_\mathrm{prev})$
\LineComment{Inner optimizer step}
\State $x \gets \texttt{BaseStep}(x, \gamma)$;\quad $n \gets n + 1$
\LineComment{Welford update within current batch; finalize and rebalance when full}
\State $c \gets c + 1$;\;\; $\delta \gets x - \mu$;\;\;
$\mu \gets \mu + \delta / c$;\;\;
$M_2 \gets M_2 + \delta \odot (x - \mu)$
\If{$c = b$}
  \State Append batch $(\mu,\; M_2 / (c - 1),\; c)$ to $\mathcal{L}$;\;\;
  reset $(\mu, M_2, c) \gets (0, 0, 0)$
  \State $\mathcal{L} \gets \texttt{Rebalance}(\mathcal{L})$
    \Comment{enforce $\le 2$ batches per merge level}
  \State $b \gets \max\bigl(\Wmin,\; 10\,\lfloor n^{1/3}\rfloor\bigr)$
    \Comment{adaptive batch size}
\EndIf
\LineComment{Reported iterate average}
\State $S \gets S + x$;\quad $k \gets k + 1$
\If{$k_\mathrm{conv} < 0$}
  \State $w \gets n_\mathrm{prev} / \gamma_\mathrm{prev}$;\quad
  $\bar x \gets (w\,\bar x_\mathrm{prev} + S) / (w + k)$
\Else
  \State $\bar x \gets S / k$
\EndIf
\LineComment{$\shR$ check: monitor stationarity within current step-size phase}
\If{$k_\mathrm{conv} < 0$ \textbf{and} batch just finalized \textbf{and} $|\mathcal{L}| \ge 3$}
  \State $\shR^\star,\; w^\star \gets \texttt{WindowedRhat}(\mathcal{L})$
  \If{$\shR^\star \le 1.1$}
    \State $W_\mathrm{chk} \gets$ sample count in best window;\;\;
    $k_\mathrm{conv} \gets n - W_\mathrm{chk}$;\;\;
    $(S, k) \gets (0, 0)$
  \EndIf
\EndIf
\LineComment{MCSE / ESS check: reduce $\gamma$ on success and start a new phase}
\If{$k_\mathrm{conv} \ge 0$ \textbf{and} $n - k_\mathrm{conv} \ge W_\mathrm{chk}$}
  \State $\bar\theta,\;\MCSE,\;\ESS \gets
    \texttt{StreamingMCSE}\bigl(\text{trailing batches with} \ge W_\mathrm{chk}\text{ samples}\bigr)$
  \State $\bar x \gets \bar\theta$
  \If{$\max_d \MCSE_d \le \epsilon$ \textbf{and} $\min_d \ESS_d \ge \ESS_{\min}$}
    \State $(\bar x_\mathrm{prev},\; n_\mathrm{prev},\; \gamma_\mathrm{prev})
      \gets (\bar\theta,\; W_\mathrm{chk},\; \gamma)$
    \State $\gamma \gets \gamma^2$;\quad
    $\epsilon \gets \gamma_\mathrm{prev}\,\epsilon$
      \Comment{quadratic step-size reduction}
    \State Reset $\mathcal{L} \gets \emptyset$,
    $(\mu, M_2, c) \gets (0, 0, 0)$,
    $b \gets \Wmin$,
    $(S, k, n) \gets (0, 0, 0)$,
    $k_\mathrm{conv} \gets -1$
  \EndIf
\EndIf
\State \Return $x,\; \bar x,\; \gamma$
\ealgc
\ealg

\subsubsection{Implementation Details}

The \texttt{Rebalance} sub-procedure enforces the geometric merging
invariant: whenever more than two batches share a level, the two oldest are
combined into a single batch at the next level using the parallel-variance
formula \cite{Chan83}, with carries propagating upward. 
The batch-size value $b \gets \max(\Wmin, 10\,\lfloor n^{1/3}\rfloor)$ ensures that
the minimum batch size grows at the MSE-optimal $n^{1/3}$ rate
\citep{Liu22}, which also ensures consistency of the batch-means
long-run variance estimator.

The diagnostic sub-procedures \texttt{WindowedRhat} and
\texttt{StreamingMCSE} operate directly on the batch summaries. For $\shR$,
following \citet{Welandawe24} we evaluate split-chain $\shR$ over several
candidate trailing windows of batches and return both the minimum $\shR$ and
the corresponding best window, which is used to identify and discard
burn-in iterates. Within each candidate window, the two halves are formed by
splitting the batch list at the boundary closest to the midpoint by sample
count, and the standard between-chain term is corrected for the resulting
unequal half-sizes. It is worth noting that we use the basic split-chain
$\shR$ rather than the rank-normalized variant of \citet{Vehtari21}, since
the latter requires access to all individual iterates.
For ESS, we generalize the equal-batch-size batch-means long-run variance estimator
\cite{Flegal10,Vats19} to the unequal batch sizes produced by merging via a
moment-matching argument. The resulting estimator is strongly consistent
when the minimum batch size grows \cite{Liu22} and tends to be conservative
(underestimating ESS), making it a good choice for convergence checking. The MCSE is
then $\sigma_d / \sqrt{\ESS_d}$, where $\sigma^2_d$ is the marginal variance
recovered from the merged batch summaries.

Finally, the reported iterate average $\bar x$ uses a ``warm-start'' Polyak
average \cite{Polyak92}: before $\shR$ convergence within a phase, $\bar x$
is a weighted mixture of the current phase's running mean and the previous
phase's converged estimate, with weight
$w_\mathrm{prev} = n_\mathrm{prev} / \gamma_\mathrm{prev}$ reflecting the
precision of the prior estimate.
Since the iterate-average error at step
size $\gamma$ decays as $O(\sqrt{\gamma / k})$, the new estimate overtakes
the prior at roughly $k = O(n_\mathrm{prev} / \gamma_\mathrm{prev})$.
Post-convergence, $\bar x$ is simply the running mean since the convergence
point. The quadratic step-size rule $\gamma \gets \gamma^2$ is derived by
matching the gradient noise at the new step size to the optimization bias
at the old \citep{Dieuleveut20}: with bias $O(\gamma)$ and iterate noise $O(\sqrt{\gamma_1 / k})$,
requiring the noise at $\gamma_1$ to match the bias at $\gamma$ yields
$\sqrt{\gamma_1} \approx \gamma$, and so $\gamma_1 = \gamma^2$. The MCSE
tolerance is rescaled by $\gamma$ at each transition to track the
corresponding bias.

\end{document}